\documentclass[preprint]{aastex}

\newcommand{\bgq}{$B_{gq}$}
\newcommand{\agq}{$A_{gq}$}
\newcommand{\Fa}{$F675W$}
\newcommand{\Fb}{$F606W$}
\newcommand{\Fc}{$F702W$}
\newcommand{\hst}{$HST$}
\newcommand{\h}{$h^{-1}_{100}$}
\slugcomment{Submitted to ApJ}

\shorttitle{WFPC2 Imaging of Quasar Environments}
\shortauthors{Finn, Impey, \& Hooper}

\begin{document}

\title{WFPC2 Imaging of Quasar Environments: A Comparison of LBQS 
and \hst \ Archive Quasars\altaffilmark{1}}

\author{Rose A. Finn, Chris D. Impey}
\affil{Steward Observatory}
\affil{933 N. Cherry Ave., Tucson, AZ  85721}
\email{rfinn@as.arizona.edu, cimpey@as.arizona.edu}
\and
\author{Eric J. Hooper}
\affil{Harvard-Smithsonian Center for Astrophysics}
\affil{60 Garden St., MS 83, Cambridge, MA  02138}
\email{ehooper@cfa.harvard.edu}
\altaffiltext{1}{Based on observations with the NASA/ESA {\it Hubble Space
Telescope}, obtained at the Space Telescope Science Institute, which is
operated by AURA, Inc.}

\begin{abstract}
We present $Hubble~Space~Telescope$ ($HST$) Wide Field Planetary Camera 2 (WFPC2) 
data on the large-scale environments of
16 $0.39 < z < 0.51$ quasars from the Large Bright Quasar Survey (LBQS).  
The LBQS quasars are representative of the radio-quiet population, and 
this is the first look at their large-scale environments.
We compare the LBQS environments with the environments of 27 
$0.15 < z < 0.55$ quasars selected from the $HST$ Archive.
The majority of the Archive quasars are 
from the PG and PKS surveys, and these quasars are more luminous on 
average than the LBQS.  By comparing the LBQS and Archive environments, 
we investigate whether previous quasar environment studies have
been biased due to studying unusually radio or optically luminous quasars.
We compare observed galaxy number counts with 
expected counts predicted from the CNOC2 field-galaxy luminosity function
in order to look for statistical excesses of galaxies around the quasars.  
We detect a significant excess around the Archive quasars 
but find no such excess around the LBQS quasars.
We calculate the amplitude of the spatial correlation function and 
find that the LBQS environments are consistent with that of the typical
galaxy while the Archive environments are
slightly less rich than Abell 0 clusters. 
We find no difference between the environments of radio-loud and radio-quiet
quasars in either sample.  However, comparison with previously
published work shows that the LBQS radio-loud quasars are in sparse environments
when compared with other radio-loud quasars, and the 
Archive radio-quiet quasars are in dense environments compared
to other radio-quiet quasars. 
The richer environments of the Archive radio-quiet quasars can not
be explained by their higher optical luminosities.  
We find a positive correlation (95\%) between radio luminosity and environment
for the radio-loud quasars.  This may explain why the LBQS radio-loud quasars,
which are less radio luminous, are in sparser environments.
\end{abstract}


\section{INTRODUCTION}
\label{introduction}
The conventional model explaining quasar activity consists of a 
supermassive black hole at the center of a galaxy.  Accretion onto the 
black hole produces emission associated with the quasar, and interactions
between the quasar host galaxy and its neighboring galaxies or intercluster
gas help to keep the accretion disk and quasar fueled.  
The importance of interactions
and mergers in triggering quasar activity is suggested by the 
high incidence of close companions 
\citep[e.g.][]{weyman78,stockton82,fg83,gehren84,malkan84,yg84,yee87,sh90}.

While the general quasar scenario is well-accepted, the origin of
the radio power from quasars is still mysterious.
\citet{blandford90} suggests that the angular momentum of a 
spinning black hole, extracted by a magnetic field, can provide the 
energy necessary to fuel radio jets.
This may imply that radio-loud quasars have black holes 
that spin more than radio-quiet 
quasars, and \citet{wc95} show that a merger event between
two black holes of relatively equal mass will result in a spinning black hole.
Such a situation might arise if two spiral galaxies merge and subsequently 
form an elliptical galaxy.  If the spiral galaxies harbor black holes of comparable
mass, then the elliptical could possess a black hole
with sufficient spin to power a radio-loud quasar.

Recent studies of the radio properties of quasars indicate
that quasars do not exhibit a bimodal distribution of radio
power, as originally thought \citep{brinkman00}.  
Radio-loudness is then more logically defined by radio luminosity,
as opposed to the ratio of radio to optical luminosity.  
Various thresholds close to log~L~=~24 in units of 
W~Hz$^{-1}$~sr$^{-1}$ 
(for $\rm H_0 = 50~km~s^{-1}~Mpc^{-1}$)
have been used to separate radio-loud and quiet quasars
\citep[e.g.][]{hooper96,bb97,goldschmidt99},
and we adopt this definition.  According to this criteria,
radio-loud quasars account for $\approx$10\% of the 
entire quasar population.

Significant progress in understanding the mechanism responsible for 
radio emission will be possible with X-ray satellites, such as XTXS. 
This telescope
will be sensitive enough to detect signatures of the black hole 
rotation in iron K-line
emission, so astronomers will be able to look for a correlation between
black hole spin and radio power.  Until then, a less direct but more 
tenable approach for understanding the radio properties of quasars
focuses on quasar host galaxies and the large-scale environments of quasars.
The first quasar host galaxy studies concluded that radio-loud 
quasars reside in elliptical
hosts and radio-quiet quasars reside in spiral galaxies 
\citep{malkan84,smith86}.
More recent work shows that while radio-loud quasars are found mostly 
in elliptical hosts, radio-quiet quasars form in both 
elliptical and spiral galaxies 
\citep[e.g.][]{bahcall97,boyce98,mclure99,sbl00}.
Quasar host studies require very precise
observations and analysis, and larger samples are needed in order to learn
what differentiates radio-loud and quiet quasars.
The high contrast between nucleus and host and the small angular scale of
the diffuse light make host studies very challenging.

Studies of the large-scale environments of quasars provide insight into
the role of environment in triggering quasar activity and can be used
to corroborate theories of radio power.
The large-scale quasar environment can also be used as
an indirect indicator of host morphology, according to the 
morphology-density relation that was first investigated by 
\citet{dressler80}.
Dressler finds that local galaxy density is linked to galaxy
type, namely, the fraction of elliptical and S0 galaxies
increases with local galaxy density.  
From the beginning, quasar environment studies have searched for statistical
excesses in galaxy counts around quasars, and we follow the same approach 
here.  A more detailed understanding of the environments of individual
quasars requires redshifts of faint galaxy companions.

This paper addresses two main questions.  First, 
have previous quasar environment studies been biased by
preferentially studying very strong radio sources or unusual optically
selected quasars?  Most studies from which the conventional picture of
quasar environments was developed employ samples drawn mainly from the
PG, PKS, 3C and 4C surveys \citep[e.g.][]{yg87,eyg91}.  
The radio surveys generally selected very luminous sources, which are
quite rare in optical samples.  Although the PG utilized selection
techniques similar to other optically selected surveys, it is 
unusual in terms of its radio properties 
\citep[e.g.][]{lafrance94,hooper96,bb97}
and optical luminosity function 
\citep[e.g.][]{wp85,kohler97,gm98}.
Quasars from the Large Bright Quasar Survey
(LBQS) are optically selected based on blue color, strong emission or 
absorption features, and a strong continuum break \citep{foltz87}.
They are representative of the
radio-quiet quasar population as a whole and are 
therefore more likely to reflect
the incidence of clustering around quasars in general.  

Second, are radio-loud quasars
located in different environments than radio-quiet quasars?
Yee and collaborators find that
environment is linked to the radio properties of quasars.
Radio-quiet quasars, while six times more likely to have a close companion
than the average field galaxy \citep{yee87}, are found in environments 
considerably less dense than those of radio-loud quasars 
(Yee \& Green 1987; Ellingson, Yee, \& Green 1991).  
Other studies with smaller samples show no evidence of a 
correlation between radio properties and environment \citep[e.g.][]{fisher96,md00}.
In apparent contradiction, a recent study of $0.5 \le z \le 0.8$ radio-loud 
quasars finds a slight but significant positive
correlation between strength of environment and radio power \citep{wold00}.
Clearly, the connection between environment and radio properties of quasars 
requires further investigation.

In this study, we present $HST$ WFPC2 data on the environments of 43 quasars spanning 
the redshift range $0.15 < z < 0.55$, and we use this information to address
the questions raised above.  
Although the field size of WFPC2 is relatively small, the superior
resolution enables an accurate separation of stars from galaxies.
This is a critical step in quantifying quasar environments.
To investigate possible biases in quasar sample selection,
we compare 16 LBQS quasars to a sample of quasars 
drawn from the Hubble Space Telescope ($HST$) Archive, 
consisting of mainly PG and PKS quasars.
The LBQS and Archive samples contain 8 and 10 radio-loud
quasars, respectively.  We compare the environmental properties 
of the radio-loud and radio-quiet quasars within each sample
and for the combined sample of LBQS and Archive quasars.

The layout of this paper is as follows.
In Section \ref{obs}, we discuss the details of the quasar samples
and observations.  In Section \ref{analysis}, we outline the procedure for
photometry and source classification and compare the observed 
galaxy counts with expected counts derived from the CNOC2 luminosity
function \citep{lin99}.
We investigate the angular distribution of any observed excess
counts (Section \ref{radial_distributions}) and estimate the amplitude of the 
spatial correlation function for each quasar (Section \ref{bgq}).  
In Section \ref{discussion}, we discuss the significance of our results
in the context of previous work, and Section \ref{conclusions} contains
a summary.
Cosmological parameters of $H_0= \rm 100~km~s^{-1}~Mpc^{-1}$, $q_0$~=~0.5, and
$\Lambda = 0$ are assumed unless otherwise noted.

\section{OBSERVATIONS AND DATA REDUCTION}
\label{obs}
\begin{figure}
\caption{Absolute magnitude, $M_V$, for the entire quasar sample versus redshift.  
The LBQS absolute magnitudes
are derived from published $B_J$ magnitudes \citep{hfc95}, 
and $M_V$ was calculated using color and $k$-corrections from
\citet{cv90}.  For the Archive sample, $M_V$ was calculated
using $V$ magnitudes from \citet{hb89} and $k$-corrections from 
\citet{cv90}.  
Radio-loud quasars are represented with filled symbols,
and radio-quiet quasars are represented with open symbols.  
The LBQS quasars are represented by triangles, and the $F606W$ and 
$F702W$ quasars are represented by squares and pentagons, respectively.}
\label{magz}
\epsscale{1.0}
\plotone{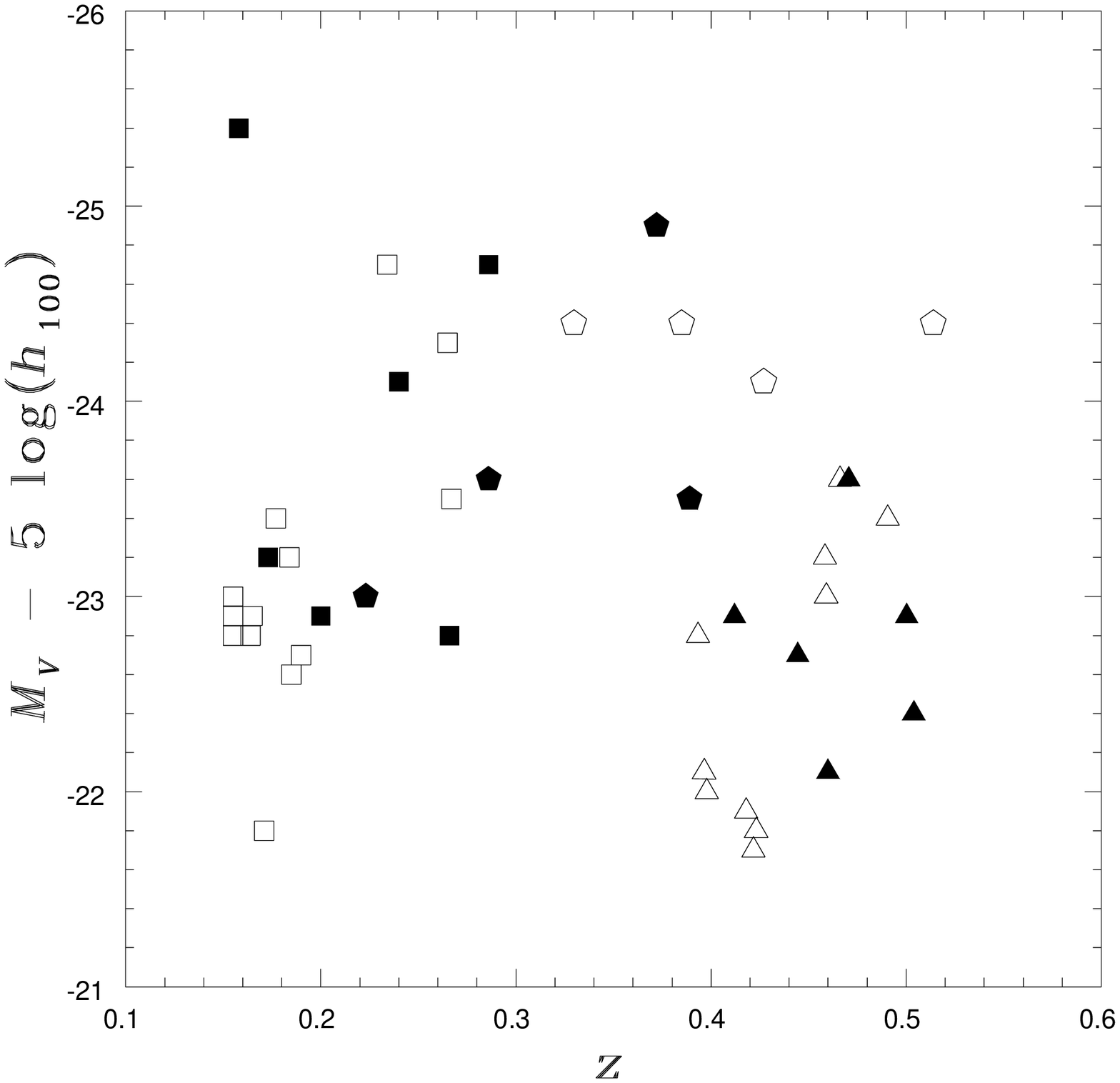}
\end{figure}

Figure \ref{magz} shows $M_V$ versus $z$ for the entire set of 
43 quasars in this environment study.  Radio-loud and 
radio-quiet quasars are
depicted with filled and open symbols, respectively.
The LBQS (Hewett, Foltz, \& Chaffee 1995) sample, shown with
triangles in Figure \ref{magz}, 
contains 16 quasars in the redshift range 
$0.39 < z < 0.504$, of which six are radio-loud.  
The remaining 10 quasars have 8.4 GHz luminosities
less than $\rm 10^{24}~W~Hz^{-1}~sr^{-1}$ (for $H_0$~=~50~km~s$^{-1}$~Mpc$^{-1}$).
The 16 LBQS quasar fields were observed with \hst \ in the $F675W$ filter
with WFPC2 during Cycle 4 with the quasar positioned on the PC, 
and an analysis of the quasar host properties
is presented by Hooper, Impey, \& Foltz (1997).  For details on 
the observations, see Table 1 and \citet{hif97}.

Details of the observations for a comparison sample of 
27 quasars drawn from the $HST$ Archive can be 
found in Table 2.  The 19 quasars observed in $F606W$
are shown with squares in Figure \ref{magz}.
These quasars have lower redshifts, with $0.15 < z < 0.29 $, and
six of the 19 are radio-loud.  All 19 $F606W$ quasars were imaged on WF3.
Studies of the host galaxies and large-scale environments of these quasars 
are presented by \citet{bahcall97} and \citet{fisher96}, respectively.
The remaining 8 quasars of the Archive sample were observed in $F702W$; 6
of the quasars were positioned on the PC and the other 2 on WF3.
The redshifts range from  $0.223 \le z \le 0.514$, and 4 out of 8 are radio-loud.  
References for the \Fc \ data are listed in Table 2.

Figure \ref{magz} demonstrates that the combined quasar environment sample is free
of the common correlation between absolute magnitude and $z$.  
A Spearman rank-correlation
test confirms this, yielding a rank-correlation coefficient of 0.04,
with an 80\% probability of getting the
same correlation from a random sample.  Furthermore, a K-S
test indicates with 84\% significance that the absolute magnitudes
for the radio-loud and radio-quiet quasars are drawn from the
same parent population.  However, a K-S test indicates only a
6\% probability that the absolute magnitudes of the Archive and
LBQS quasars are drawn from the same parent population, meaning 
that on average the Archive quasars are more luminous than the LBQS quasars.

For all quasar data, calibrated images are retrieved from the $HST$ Archive, 
and cosmic ray rejection is achieved with the STSDAS combine routine with 
the ``crreject'' option set.  
Magnitude zeropoints for the three $HST$ filters are from Table 9 in
\citet{holtzman95}.  The zero points are adjusted for differences in gain 
and are increased by 0.1 to convert to infinite aperture magnitudes. 

\section{ANALYSIS}
\label{analysis}
\subsection{Photometry and Geometry}
SExtractor is used for photometry and source classification
\citep{ba96}.  For the wide-field cameras,
a detection threshold of 1.5$\sigma$ per pixel
is used with a minimum object size of 32 pixels.   
The PC images are binned by 2$\times$2 to improve signal-to-noise, and 
a detection threshold of 1$\sigma$ per pixel with a minimum
object size of 32 pixels is used.
Total magnitudes are determined using the ``mag-auto'' algorithm.
A tophat 5$\times$5 convolution kernel is used for 
both the PC and WF data. 

\begin{figure}
\caption{SExtractor classifier index (1=unresolved) versus magnitude
for the entire LBQS $F675W$ dataset.  Triangles represent WF data; 
open circles represent PC data.  
A classifier index of 1 represents a star and 0 a galaxy; 
objects with indices below 
0.4 are considered galaxies.
PC data with indices falling between 0.5 and 0.95 correspond to the quasars.  
}
\label{sextractor}
\epsscale{1.0}
\plotone{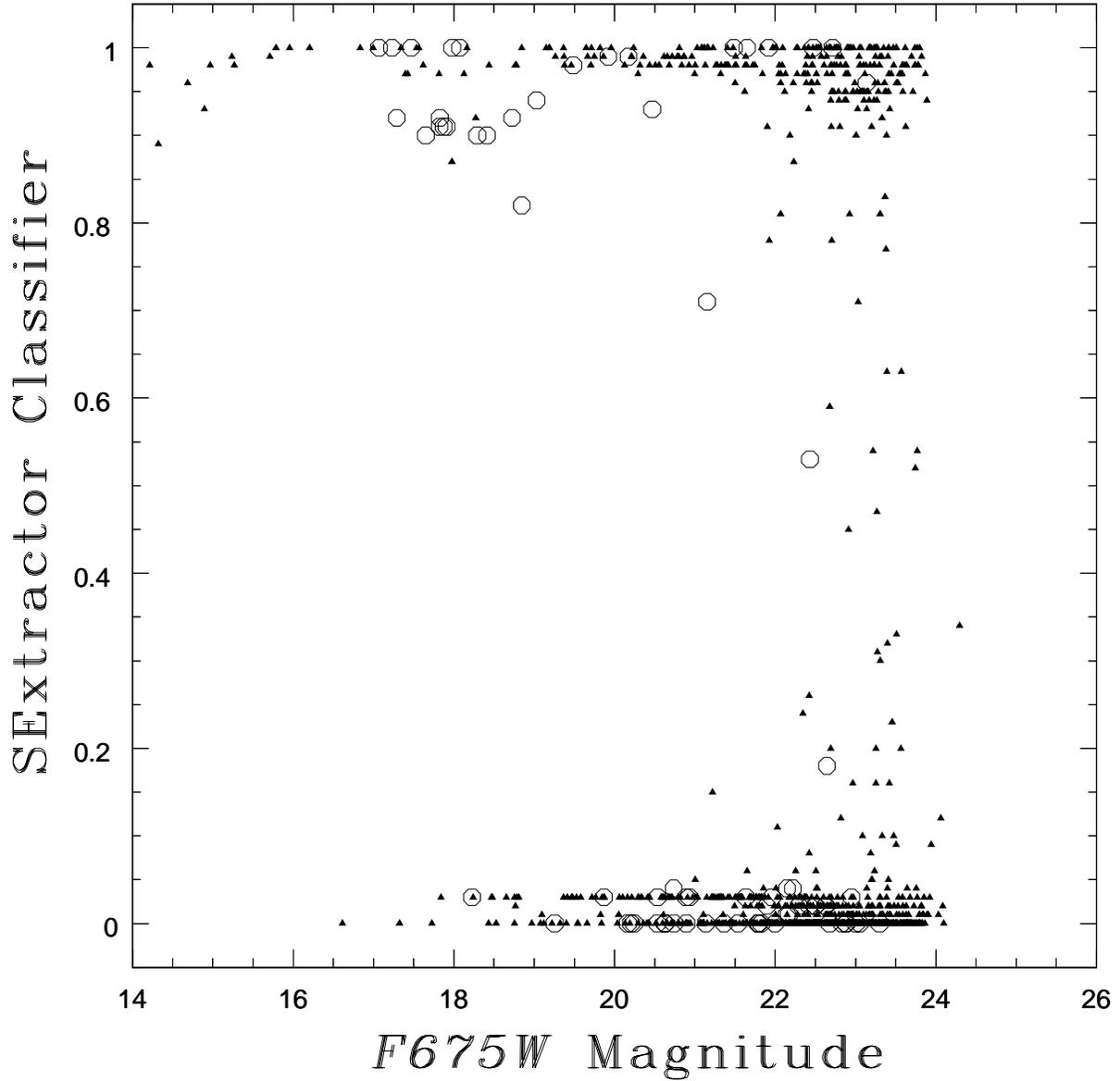}
\end{figure}

SExtractor does a remarkably good job at classifying objects in $HST$ data.
Figure \ref{sextractor} shows the SExtractor classification index versus magnitude
for all of the \Fa \ LBQS fields, where a classification index of 1 indicates
a star or unresolved source and 0 a galaxy.  We 
achieve a clean star/galaxy separation down to
$m_{F675W} = 22$ for the LBQS data.  
In Figure \ref{sextractor}, 
the bright ($m < 20$) objects in the LBQS PC fields with classifier indices
between 0.5 and 0.95 correspond to the quasars,
showing that a number of them are marginally resolved.
The results of the star/galaxy separation are comparable for the 
Archive fields.
We use a SExtractor classification index of 0.4 as our cut-off for galaxies.
Note that the analysis is not sensitive to this value, since for 
$m_{F675W} < 22$, the distribution of classification indices is bimodal,
with very few objects having indices greater than 0.1 and less than 0.9.

\begin{figure}
\caption{(a) Input (solid) distribution of artificial galaxies
with the distribution of galaxies recovered by SExtractor (dotted) for a
1200~second $F675W$ PC image. 
The data are binned $2\times2$. 
SExtractor detection parameters include a threshold of 1.0$\sigma$ per 
pixel and a minimum object area of 32 pixels.  
(b) Completeness versus magnitude.
No galaxies with $m_{F675W} > 22$ are used in this analysis.}
\label{completeness}
\epsscale{1.0}
\plotone{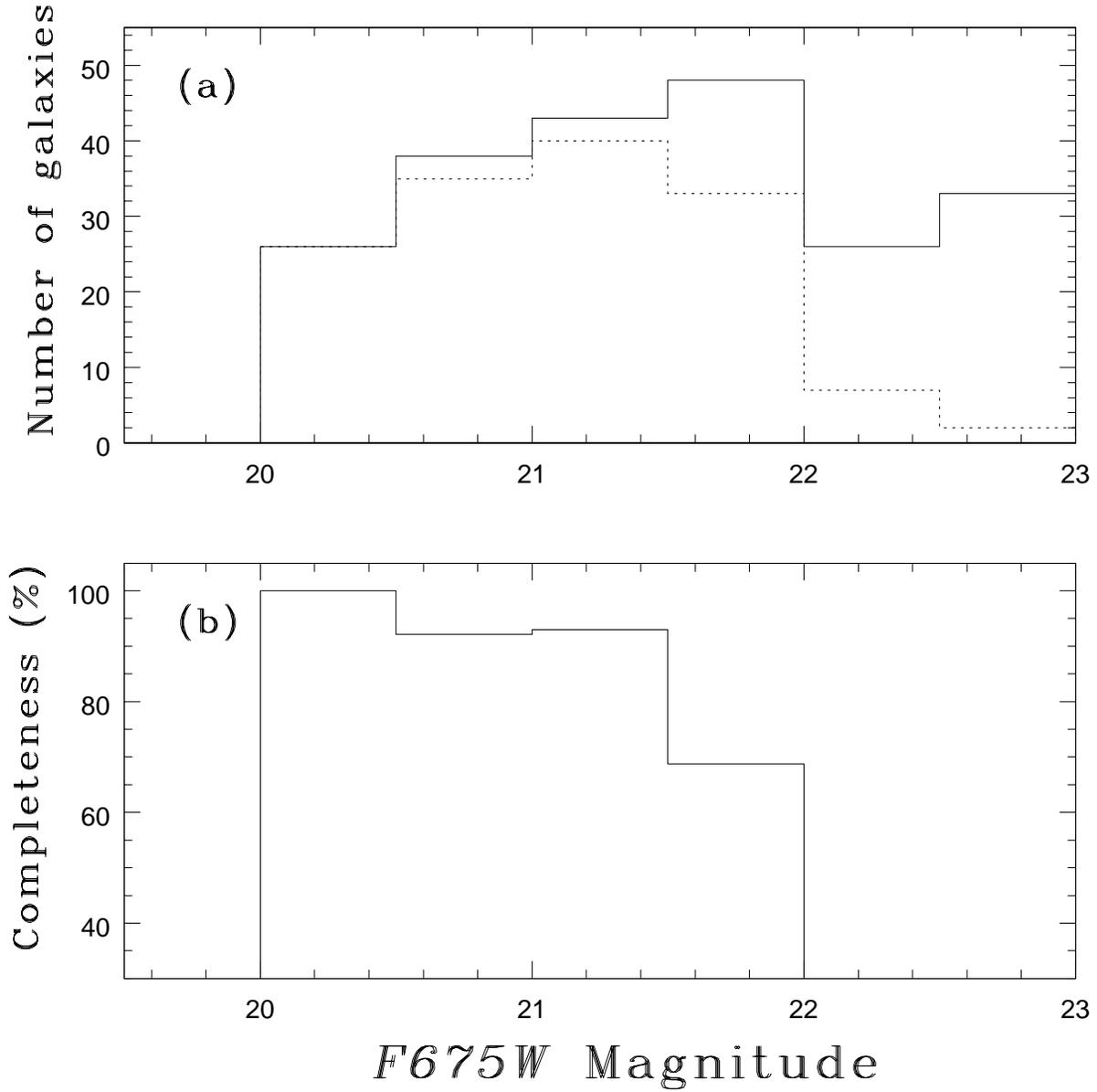}
\end{figure}

We determine the completeness of the data by adding artificial
galaxies to the final images using the IRAF ``artdata'' package.
Parameters for the artificial galaxies are based on data from 
Griffiths et al. (1994a, 1994b)
and include: 1\arcsec \ for
the maximum half-light radius of a $m_{F675W} =21$ elliptical galaxy,
a value of one for the ratio of elliptical to spiral half-light radii, and 
0.3 for the fraction of ellipticals.
For the \Fa \ data, we place artificial galaxies with $20 \le m_{F675W} \le 23$
on the PC and WF images.  We then estimate completeness based
on how many of the artificial galaxies were recovered by SExtractor. 
In the 1200 second \Fa \ data, 100\% of $21.5 \le m_{F675W} < 22$ galaxies
are recovered on the wide-field images.  The results for the
PC images are shown in Figure \ref{completeness}.  The completeness
drops to $\sim$95\% for $20.5 < m < 21.5$ galaxies and falls below 80\% 
for $21.5 < m < 22$ galaxies.  To assess the significance of this
incompleteness, we compute the expected galaxy 
counts per field (as discussed in Section \ref{num_counts}).  
We find that the completeness-corrected counts 
do not differ significantly from the
uncorrected counts down to $m_{F675W} = 22$.
We therefore use $m_{F675W} = 22$ as the faint magnitude cut for the 
\Fa \ data and do not apply a completeness correction.    
The Archive data suffer from a similar level of incompleteness,
so again we limit our analysis to $m_{F606W} \le 22$ and $m_{F702W} \le 22$
galaxies for the Archive quasars and make no other adjustments for
completeness.

Before proceeding, we need to comment on the geometry of the images.
First, all the LBQS quasars are centrally positioned on the PC.  
Six of the Archive 
quasars are also positioned on the PC, but the remaining 
21 are located on WF3.
Our results could be affected by the asymmetric 
geometry of WFPC2.  For example, if a quasar positioned on the PC is 
located at the edge of a cluster, as found in a recent study of $1.0 < z < 1.6$
radio-loud quasars \citep{sg99}, 
we might miss the cluster entirely.  
The inferred environmental density will then be underestimated.  
We will be better able to comment on 
the effect of WFPC2 geometry when we analyze our wider-field, ground-based 
images of the LBQS quasars (see Section \ref{conclusions}).
Another issue that complicates any measure of local density
around the quasars is the relatively
small field size of WFPC2.  At a redshift of 0.3, the 2.5\arcmin \
field of WFPC2 corresponds to a projected size of 0.58 \h~Mpc, whereas
the expected diameter of a group is $\sim$1~Mpc \citep[e.g.][]{zm98}.  
Therefore, for the lower-redshift quasars especially,
we are sampling a limited volume around the quasar.  Furthermore, since
a group or cluster in which the quasar is located could fill the entire
WFPC2 field, we cannot use the periphery of the images to 
estimate expected field-galaxy counts.
These issues are addressed in the next section, with a 
direct comparison with wide-field galaxy counts.

\subsection{Comparison of Galaxy Number Counts with Surveys}
\label{num_counts}
Our first step in looking for excess galaxies around the quasars is to compare
the observed galaxy number counts with published surveys.
This allows us to make a reliable estimate of expected galaxy counts even
though a group or cluster in which a quasar is located may fill the entire
WFPC2 field.  Unfortunately, we
find no deep galaxy surveys conducted in the three $HST$ filters of interest.
(The Medium Deep Survey includes \Fb \ data, but number counts have not 
been published yet.)
Converting number counts from other filters requires detailed knowledge of
the galaxy populations involved, which leads us to the CNOC2 dataset.
\citet{lin99} have calculated the galaxy luminosity function based
on CNOC2 data, and they parameterize the evolution of the 
luminosity function with redshift for both the $B$ and $R_C$ filters.
\citet{lin99} classify galaxy SEDs using the \citet{cww80} templates, and  
knowing the percentage of different galaxy types 
allows one to transform the luminosity function parameters from $R_C$ to any band, 
in our case the $HST$ $F606W$, $F675W$ and $F702W$ filters. 
To recover the integrated luminosity function for the $HST$ filters, 
we use eq. (23) from \citet{lin99} and
the $R_C$ luminosity function parameters.  We obtain the absolute $R_C$ magnitudes
from observed $HST$ magnitudes by calculating the $k$- and color-corrections from 
the Coleman, Wu \& Weedman SEDs.  The zero point of each filter 
is determined from the spectrum of Vega.
We integrate the luminosity function between $0 < z < 1$; integrating beyond
$z = 1$ does not affect the counts below $m = 22$, but predicted counts at fainter 
magnitudes are affected. 

\begin{figure}
\caption{Observed galaxy counts per square degree (triangles) with
the integrated luminosity function (solid line) derived from CNOC2 galaxy survey 
\citep{lin99}.  Panels (a), (b) and (c) show the results for the 
CNOC2 $R_c$, $V$ and Medium Deep Survey $F555W$ \citep{casertano95}
counts, and the
integrated luminosity function matches survey data well.  Panels (d), (e) and (f)
show the galaxy counts derived from WFPC2 images of quasar fields.
The integrated luminosity function matches the $F675W$ galaxy counts, 
but the quasars imaged 
in the $F606W$ and $F702W$ filters show excess counts.  
Error bars are 1$\sigma$ Poisson errors.}
\label{intlf}
\epsscale{1.0}
\plotone{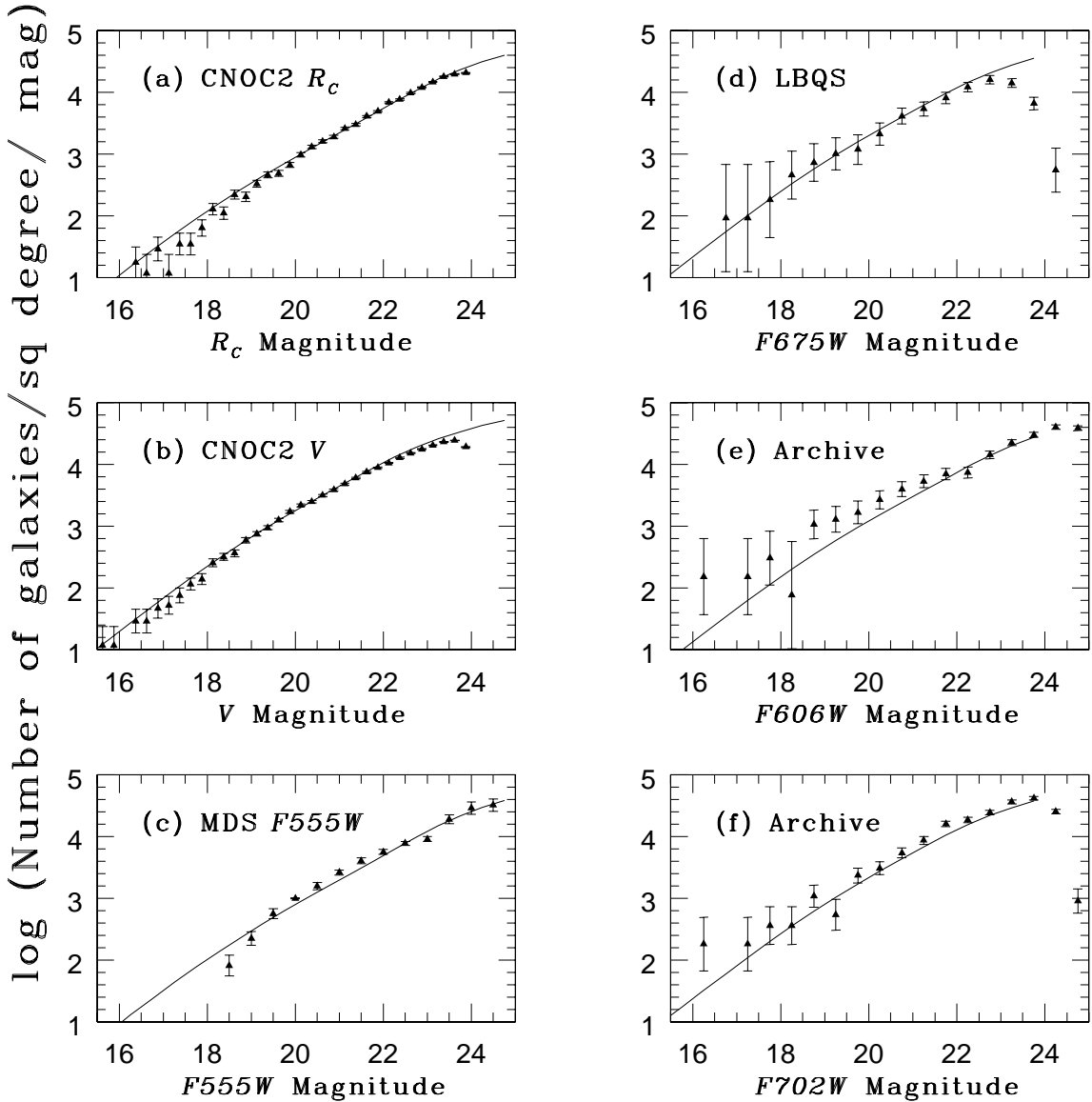}
\end{figure}

As a test of the method, we calculate the integrated luminosity function for 
$V$, $R_C$ and $F555W$ filters.  
Panels (a) through (c) of Figure \ref{intlf} show 
that the integrated luminosity function 
reproduces the CNOC2 $V$ and $R_C$ counts (H. Lin, private communication) 
as well as 
the Medium Deep Survey F555W counts \citep{casertano95}.
The CNOC2 integrated luminosity function overpredicts the number 
of galaxies brighter than 19th magnitude.
In terms of galaxies expected
in a WFPC2 field with $m_{R_C} < 22$, the integrated luminosity function 
predicts 32.6 galaxies per field while the 
CNOC2 counts give 31.5 galaxies per field.  In the $V$ band, an average of 16.2 
galaxies are expected and 16.0 are observed.  This means that in our $F675W$ and 
$F702W$ quasar fields, the expected field galaxy counts will be systematically 
high by $\approx$1 galaxy per field.  For the \Fb data, which are closer to $V$, 
the expected field galaxy counts will be systematically high be
$< 0.5$ galaxy per field.

Our next step in looking for excess galaxies is to use the CNOC2 
integrated luminosity function to 
estimate the expected galaxy counts in the quasar fields.
Implicit in this comparison is the assumption that the magnitudes
derived from SExtractor are equivalent to the CNOC2 galaxy
magnitudes.  The CNOC2 galaxy magnitudes are determined using
the PPP software \citep{yee91}.  To check the validity of this assumption,
H. Lin ran SExtractor and PPP on the same data.  A comparison
of the SExtractor and PPP total magnitudes
shows $\sim$0.2 magnitudes of scatter but no systematic offset
(H. Lin, private communication).  We therefore proceed in 
using the CNOC2 integrated luminosity function to predict 
the expected galaxy counts in the quasar fields.

The CNOC2 luminosity function is integrated for the 
$HST$  $F675W$, $F606W$ and $F702W$ filters, and the results are 
plotted with the galaxy counts from the quasar fields in panels
(d) through (f) of Figure \ref{intlf}.
The integrated $F675W$ luminosity function reproduces 
the observed galaxy counts from the LBQS fields within the errors
down to $m \approx 23$.
The $F606W$ Archive galaxy counts show a clear excess at magnitudes brighter than
21, but the integrated luminosity function matches the faint galaxy 
counts well.    
The $F702W$ Archive galaxy counts show a slight excess above the 
predicted counts at almost all magnitudes.

To assess the significance of these findings, we translate 
from galaxies per square degree per magnitude
into the average expected and observed counts per field.
To determine the expected galaxy counts per field, we sum the 
integrated luminosity function
between $14 < m < 22$ and multiply by the area of WFPC2 in square degrees.
This pushes slightly beyond the spectroscopic completeness
level of $R_C = 21.5$ for the CNOC2 survey.  
The observed, expected, and excess counts for the individual LBQS fields are listed
in columns 5, 6 and 7 of Table 3.  The same quantities are listed 
for the \Fb \ and \Fc \ Archive fields in Tables 4 and 5.  
We find that the LBQS quasars
have an average of 16.1 galaxies per field, $1.8 \pm 1.4$ galaxies less than 
expected.
The Archive \Fb \ and \Fc \ quasars have an average of 16.4 and 
26.5 galaxies per field,
$5.4 \pm 1.0$ and $7.0 \pm 2.0$ galaxies more than expected, respectively.  
The error in the excess counts
is determined by propagating the error in expected counts.  The error
in expected counts is taken to be 
1.3 times the Poisson error, where the extra factor of 1.3 
accounts for variations associated
with large-scale structure, as observed by \citet{ygs86}.  
We see no significant excess associated with the LBQS quasars.  The Archive
quasars, in comparison, have very significant excesses.
The average excess of 5.4 galaxies per field
associated with the \Fb \ Archive quasars is a total that is
slightly more than 5$\sigma$ above the noise associated
with the expected galaxy counts.  The average excess of 7.0 galaxies
per field detected in the \Fc \ Archive images is a 3$\sigma$ result overall.

Previous quasar environment studies have limited the analysis to galaxies
with $m < m_* + 2.5$ \citep{yg87} or $m_*-1 < m < m_* +2$
\citep{fisher96,md00}, 
where $m_*$ is the apparent magnitude of the
knee of the luminosity function at the redshift of each quasar.  This
acts to reduce the contamination from field galaxies.
Since all of the \Fb \ Archive fields are common to the Fisher et al.
dataset, we apply a magnitude slice of
$m_* - 1 < m < m_* + 2$ to facilitate comparison.  
For consistency with the estimation of
expected counts, we use the value of $m_*$ 
derived from CNOC2.  Values of $m_*$
are listed in column 10 of Tables 3, 4, and 5 for the LBQS, \Fb, 
and \Fc \ Archive quasars, respectively.  In addition, the
observed, expected, and excess number of galaxies are listed in
columns 11, 12, and 13.  Applying this limited magnitude
slice leaves the average observed excess for the LBQS fields
unchanged: $-1.8 \pm 1.4$ becomes $-1.7\pm 1.4$ galaxies.   
The negative sense of the average for the LBQS sample is not
significant. 
The excess 
for the \Fb \ Archive fields drops from $5.4 \pm 1.0$ to 
$3.7 \pm 0.7$ galaxies per field.
The excess for the \Fc \ Archive fields drops from 
$7.0 \pm 2.0$ to $5.8 \pm 1.8$ galaxies per field.
This restricted magnitude slice does not change the significance of the observed
excesses.  

\subsection{Radial Distribution of Galaxies}
\label{radial_distributions}
\begin{figure}
\caption{Average number of excess galaxies per field 
versus angular distance from the quasar
for the (a) Archive and (b) LBQS samples.  
Error bars are 1$\sigma$ Poisson errors.
The Archive sample shows an excess above background, but
the LBQS dataset shows no significant excess.}
\label{rankang}
\epsscale{1.0}
\plotone{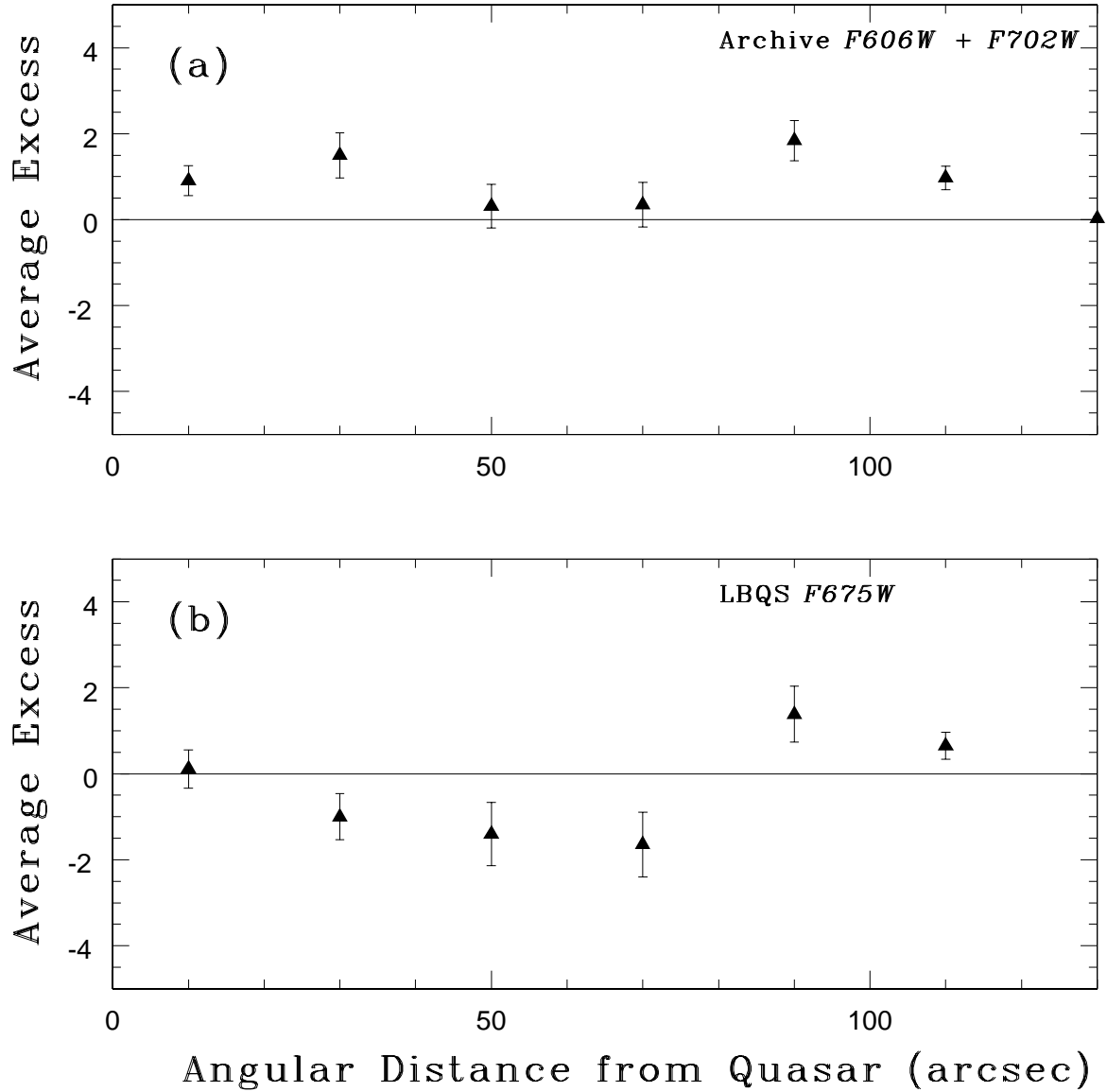}
\end{figure}

\begin{figure}
\caption{Average number of excess galaxies per field versus 
projected distance to quasar
for the (a) Archive and (b) LBQS quasar samples.  
The Archive quasars show a centrally-concentrated excess above background
out to 300 $h^{-1}_{100}$~kpc; the LBQS quasars show no excess.
Error bars are 1$\sigma$ Poisson errors.}
\label{radial_lstar}
\epsscale{1.0}
\plotone{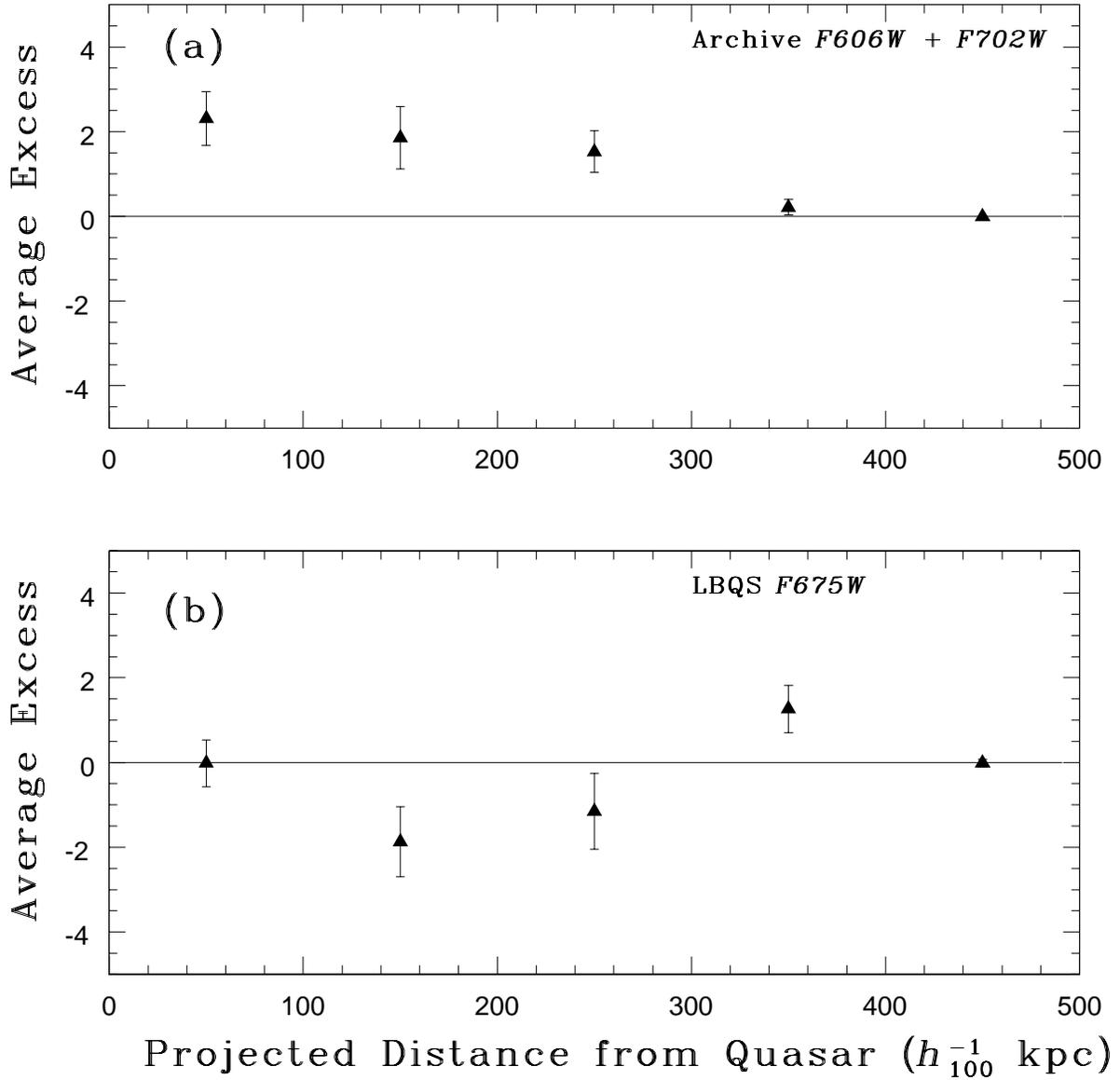}
\end{figure}

We detect an excess of galaxies above the 
CNOC2-derived field estimate in the Archive data.
We observe no significant excess in counts in the LBQS fields.
If we assume that the excess galaxies observed in the 
Archive fields are associated with the quasar and are 
located at the quasar redshift, then we might expect the excess
counts to originate from regions close to the quasar,
as seen in other studies 
\citep[e.g.][]{yg87,eyg91,sbm95,fisher96,hg98}.
Figure \ref{rankang} shows the average excess number of 
galaxies per field versus
angular distance from the quasar for the Archive and LBQS samples.
The Archive fields show an excess 
with no obvious radial dependence, and the LBQS show no excess.
Since the Archive quasars span a large range in redshift,
a given angular distance corresponds to a very different
projected distance for the lowest and highest redshift 
Archive quasars.  This could dilute an observed radial
gradient of galaxies.  Therefore, in Figure 
\ref{radial_lstar} we plot the spatial distribution 
of excess galaxies.  Here we
assume that all the galaxies in the field are at the quasar redshift and 
convert angular distance to projected physical distance.  
Counts are binned in equal-width annuli of 100~\h~kpc.
This comparison shows 
a highly significant excess of galaxies out to 400~\h~kpc for the Archive quasars 
and no excess for the LBQS quasars.  
The difference between Figures \ref{rankang} and \ref{radial_lstar}
for the Archive data
supports the assumption that the excess galaxies are associated with the quasar.
The significance of the drop-off in excess counts at radii larger than 300~\h~kpc
is difficult to assess due to the limited field size of WFPC2.

In looking for a radial gradient in galaxy counts around a given
quasar, we are assuming that the quasar is at the center of the local mass
concentration.  A recent study of higher-redshift quasars shows that this
is not necessarily the case \citep{sg99}.  
Unfortunately, the WFPC2 field is not big enough to explore this scenario.
The possibility that quasars are not centered 
in their groups or clusters will be addressed with existing ground-based data for
the LBQS sample, and these results will be presented in a future paper.

\subsection{Amplitude of the Spatial Correlation Function}
\label{bgq}
One common way to quantify the strength of clustering around quasars is
to calculate the amplitudes of the angular and spatial correlation functions.
For a given cluster around a quasar, the amplitude of the angular 
correlation function, $A_{gq}$,
will decrease with increasing quasar redshift.  The amplitude 
of the spatial correlation function, $B_{gq}$, takes redshift
into account by projecting
angular distances into physical distances.  Therefore,
$B_{gq}$ is a better way to compare our samples.  
However, \agq \ is the parameter that is measured observationally.
\citet{ls79} derived the 
relation between \agq \ and \bgq, and
a brief summary of their work is given below.

The angular correlation function, $\omega(\theta)$, is defined by
\begin{equation}
\label{ang_corr}
n(\theta) \ d\Omega = n_{exp}\ (1 + \omega(\theta)) \ d\Omega,
\end{equation}
where $n(\theta) \ and \ n_{exp}$ are the surface densities of observed and
expected galaxies, respectively, and $d\Omega$ is surface area.
The spatial correlation function, $\xi(r)$, is defined as the excess 
number of galaxies at a distance $r$ from an object 
in a volume element $dV$,  
\begin{equation}
\label{corr_func}
\label{spatial_corr}
\rho(r)\ dV = \rho_{exp}\ (1 + \xi(r)) \ dV.
\end{equation}  
The quantities $\rho(r) dV$ and $\rho_{exp} dV$ are the number of galaxies observed and
expected in volume element $dV$, respectively.
\citet{ls79} show that $\xi(r)=B_{gq} \ r^{-\gamma}$ 
implies $\omega(\theta) = A_{gq} \ \theta^{-(\gamma-1)}$.
In deriving the relationship between \agq \ and \bgq, 
\citet{ls79} find
\begin{equation}
\label{bgq_agq}
B_{gq} = \frac{A_{gq} \ n_{exp}}{\phi(m_o,z) \ {(\frac{D}{1+z})}^{3-\gamma} \ I_\gamma},
\end{equation}
where $\phi(m_o,z)$ is the luminosity function integrated at the quasar 
redshift down to limiting magnitude $m_o$, $D$ is the effective 
distance (making $D/(1+z)$ the angular diameter distance, $D_A$), and $I_\gamma$ 
is the integration constant that arises from 
integrating the volume in a cone  
corresponding to surface area $d\Omega$.

In practice, in order to calculate \bgq, we must first calculate \agq.  
Substituting the functional form of $\omega(\theta)$ from \citet{ls79} into 
equation \ref{ang_corr}, integrating 
$d\Omega$ over the image area and solving for $A_{gq}$, we find
\begin{equation}
\label{agq_final}
A_{gq} = \frac{N_{obs} -  N_{exp}}{n_{exp} \ \sum \theta^{-(\gamma-1)} \ \Delta\Omega}.
\end{equation} 
The final expression for \bgq, obtained by substituting equation 
\ref{agq_final} into equation \ref{bgq_agq}, is
\begin{equation}
B_{gq} = \frac{N_{obs} - N_{exp}}{\phi(m_o,z) \ D_A^{(3-\gamma)} \ I_\gamma \ 
\sum \theta^{-(1-\gamma)} \ \Delta\Omega}.
\end{equation}
If we consider only the Poisson error associated with $N_{exp}$ and $\phi(m_o,z) 
D_A^3$, the uncertainty in \bgq \ is
\begin{equation}
\label{bgq_err}
\frac{\Delta B_{gq}}{B_{qg}} = 
\frac{1}{N_{obs}- N_{exp}}
\sqrt{1.3^2 N_{exp} + \frac{(N_{obs}-N_{exp})^2}{\phi(m_o,z) \ D_A^3}}.
\end{equation}

In calculating \bgq, we use $\gamma = 1.77$, the power law index derived 
from local studies
of the galaxy-galaxy covariance function \citep{sp78}. 
\citet{yg87} demonstrate that $\gamma = 1.77$ is appropriate to use
for $z < 0.6$ quasars. 
This makes the units of \bgq \ Mpc$^{1.77}$.
We evaluate $\sum \theta^{-\gamma} \Delta \Omega$ by summing 
$\Delta\Omega$ in concentric, quasar-centered
annuli from zero radius to the most distant point in the field.  
The field counts, $N_{exp}$, are derived from the CNOC2 luminosity
function
\citep{lin99} as described in Section \ref{num_counts}. 
We also use the CNOC2 luminosity function for calculating $\phi(m_o,z)$. 

The results of the \bgq \ calculations are listed in Tables 3, 4, and 5
for the LBQS, \Fb, and \Fc \ Archive samples.
The layouts of Tables 3, 4, and 5 are identical.  Columns 1 and 2 give
the quasar name and redshift.  The angular distance from the
quasar to the furthest corner of the WFPC2 field is listed in
column 3, and the physical distance that this angle corresponds
to at the quasar redshift, $r_{max}$, is listed in column 4.  Columns 5 through
9 correspond to all galaxies within the magnitude range
$14 < m < 22$.  Columns 5 and 6 give the total number of observed
and expected galaxies per WFPC2 field.  The error in the 
expected counts are 1.3 times the Poisson error, as 
described in Section \ref{num_counts}.  Column 7 gives
the excess number of galaxies per field, and column
8 expresses this excess in terms of $\delta$, the ratio
of the excess counts to the expected counts.
Column 9 lists the calculated values of \bgq, with the errors derived
according to equation \ref{bgq_err}.  Columns 10 through
15 refer to only those galaxies with $m_* -1 < m < m_* +2$.
Column 10 gives the value of $m_*$ for each quasar,
and columns 11 through 15 are analogous to
columns 5 through 9, but for the narrower magnitude range.

The average \bgq \ for the LBQS sample is $-16 \pm 13$~(\h~Mpc)$^{1.77}$.
As with the average number counts for the LBQS sample, the negative sense
of the average \bgq \ value is not significant.  
The average values of \bgq \ for the \Fb \ and \Fc \ Archive samples 
are  $59 \pm 11$ and $58 \pm 18$~(\h~Mpc)$^{1.77}$.
Considering $m_* -1 < m < m_* +2$ galaxies only leaves 
the average value of B$_{gq}$ unchanged for the LBQS sample.  The
averages for the \Fa \ and \Fc \  Archive samples are 
$60 \pm 10$ and $49 \pm 18$~(\h~Mpc)$^{1.77}$.
For comparison, \citet{dp83} find that the average
value for the amplitude of the galaxy-galaxy spatial correlation function, 
$B_{gg}$, is 20~(\h~Mpc$)^{1.77}$, and
\citet{ls79} find that clusters with Abell richness classes 0 and
1 have $B_{gg}$ values of 90 and 250~(\h~Mpc$)^{1.77}$, respectively.
The LBQS quasar environments are consistent with that of a typical
galaxy, while the average Archive quasar environment
is slightly less rich than an Abell 0 cluster.
We note that the relative values and significance levels of 
\bgq \ for the various samples match the simple estimates of 
galaxy excess with respect to published surveys, as expected.

Since our quasar sample spans a wide range of redshifts, measuring
\bgq \ out to the edge of the field could introduce systematic differences
in the results calculated for low and high-redshift quasars.
To make the measurements more uniform, we recalculate \bgq \ for galaxies within
a projected distance of 200~\h~kpc from the quasar.  
This corresponds to the largest physical size
imaged by WFPC2 for the lowest redshift quasar in the Archive sample.
For $14 < m < 22$ galaxies only, the average value of \bgq \ 
for the LBQS sample is $\rm 0 \pm 11$~(\h~Mpc)$^{1.77}$.  
The average values for the 
\Fa \ and \Fc \ Archive samples are $\rm 53 \pm 
10 \ and \ 65 \pm 16$~(\h~Mpc)$^{1.77}$.  These values do not change 
significantly when considering only galaxies
with $m_* - 1 < m < m_* +2$.  
Thus, the distance at which \bgq \ is measured does not affect the average
\bgq \ values significantly.  
Unless otherwise stated, we will refer
to the \bgq \ values calculated for $m_* - 1 < m < m_* +2$ galaxies using the
entire WFPC2 field for the remainder of the paper.

\begin{figure}
\caption{$B_{gq}$ versus redshift for the entire sample of quasars.
Radio-loud quasars are represented with filled symbols,
and radio-quiet quasars are represented with open symbols.  
The LBQS quasars are represented
by triangles, and the $F606W$ and $F702W$ quasars are 
represented by squares and pentagons, respectively.  There is no evidence
for a radio dichotomy, and $B_{gq}$ values for the LBQS sample 
lie systematically below the values of the Archive sample.}
\label{bgqz}
\epsscale{1.0}
\plotone{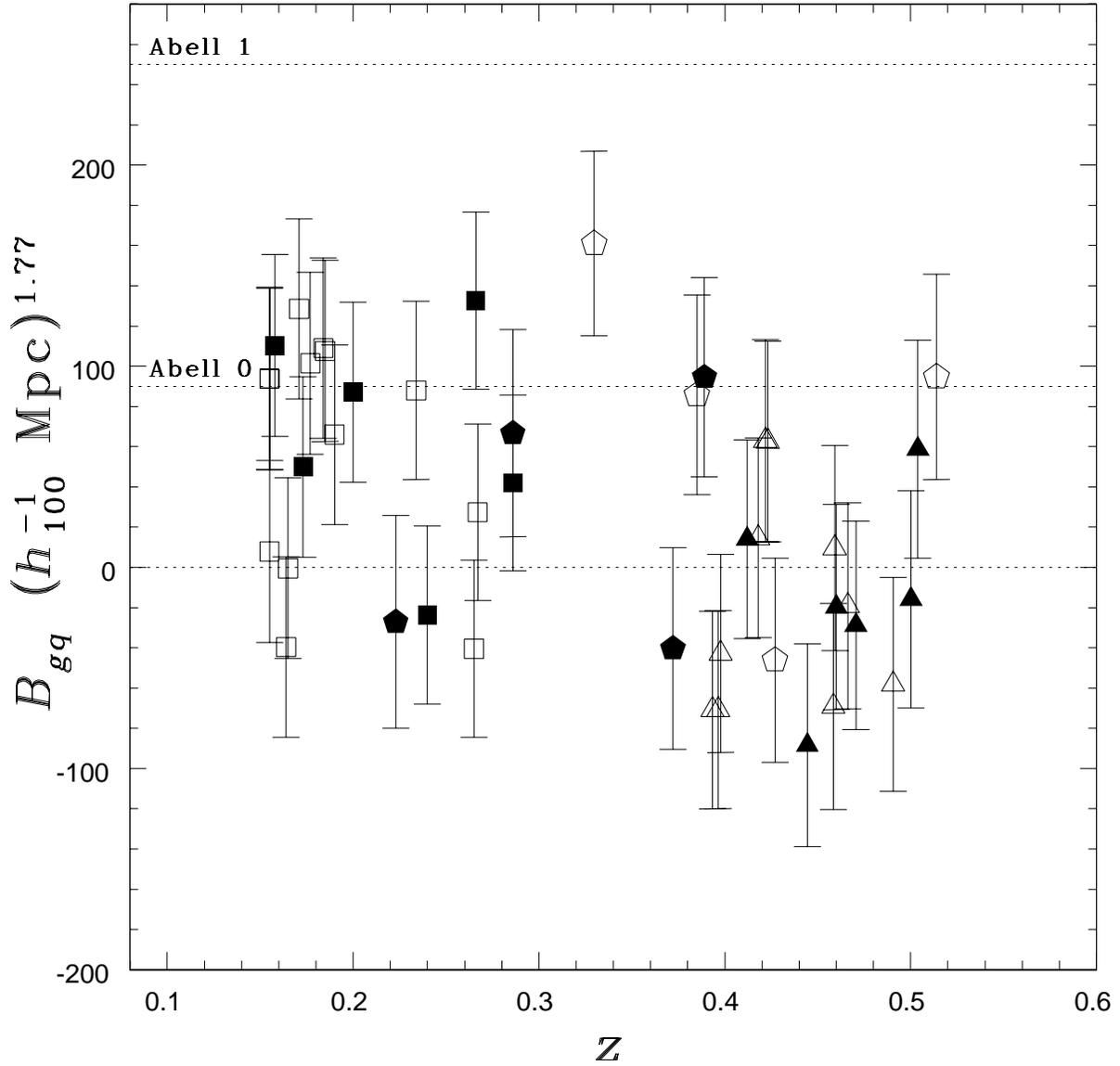}
\end{figure}

Figure \ref{bgqz} shows a plot of \bgq \ versus redshift, and 
three interesting points emerge from this figure.  
First, we find no apparent radio dichotomy.  A K-S test indicates  
that the radio-loud and radio-quiet samples 
are drawn from the same population with 58\% confidence.
Similar results are obtained for the three subsamples individually;
for the LBQS, \Fb, and \Fc Archive fields, a K-S test indicates a 
70\%, 56\% and 53\% probability that the radio-loud and radio-quiet 
subsamples are drawn from the same distribution of \bgq \ values.
The similarity of radio-loud and radio-quiet environments is 
consistent with results of \citet{fisher96} and \citet{md00} 
but is not consistent with the 
findings of \citet{eyg91}.  We will discuss this point
in more detail in Section \ref{discussion}.

\begin{figure}
\caption{$B_{gq}$ versus redshift, with Archive $B_{gq}$ values recalculated 
using only
$m_* - 1 < m < m_* + 1.4$ galaxies to mimic the completeness of the LBQS images.  
Radio-loud quasars are represented with filled symbols,
and radio-quiet quasars are represented with open symbols.  
The LBQS quasars are represented
by triangles, and the $F606W$ and $F702W$ quasars are 
represented by squares and pentagons, respectively.  Again,
the LBQS sample lies systematically below the values of the Archive sample.}
\label{limitmag_z}
\epsscale{1.0}
\plotone{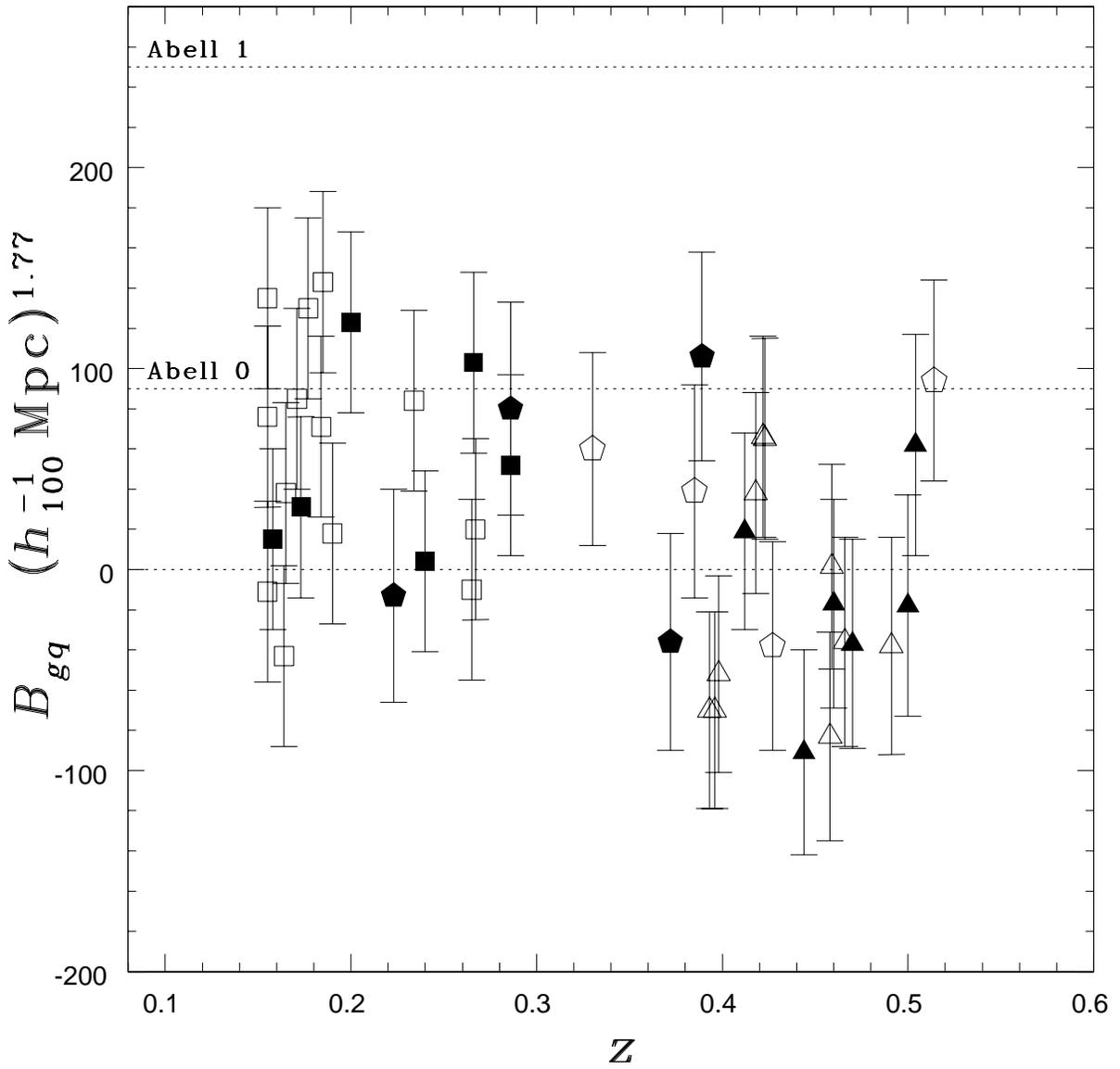}
\end{figure}

The second point to draw from Figure \ref{bgqz} is that the values of
$B_{gq}$ for the LBQS sample lie systematically below those of 
the Archive quasars.  
The significance of this difference is 99\%, as determined by the K-S test.
This is consistent with the differences found between the LBQS and
Archive samples based on number counts, described in Section \ref{num_counts}.
Is something different about the LBQS 
environments, or is there some systematic error associated with the 
LBQS data or analysis?  One possible systematic is the 
shorter exposure times of the LBQS data.  Therefore, we might not be sampling far
enough down the luminosity function to pick up companions.
By applying a limiting magnitude cut of $m_{F675W} = 22$, we sample to an
average depth of $m_* + 1.4$ in the LBQS data.
To see how this shallower depth affects the inferred environmental density, 
we recalculate \bgq \ for the Archive fields using only $m_* -1 < m < m_* +1.4$ 
galaxies.
The resulting \bgq \ values are plotted versus quasar redshift in 
Figure \ref{limitmag_z}.  A K-S test again indicates only a 0.6\% probability that
the \bgq \ values for LBQS and Archive quasars are drawn from the 
same parent distribution.  We must then conclude that the LBQS quasars 
are located in environments less dense than the Archive quasars.

The final conclusion drawn from Figures \ref{bgqz} and \ref{limitmag_z}
is that we do not see an increase in \bgq \ with increasing redshift out to 
$z = 0.5$.  \citet{hl91} see a significant 
enhancement in the environments of 
radio galaxies with moderate to high radio power by $z \sim 0.5$, 
and \citet{yg87} find evidence for a strong
increase in the density of environments of radio-loud quasar at $z > 0.6$.
A Spearman rank test performed on the data in Figure \ref{limitmag_z} 
indicates a negative correlation between \bgq \ and $z$
at the 99\% confidence level.  However, this correlation disappears when
considering the Archive data only, with a Spearman rank
test indicating an 86\% probability that no correlation exists.  
Therefore, the correlation of
\bgq \ with $z$ is due to the systematically lower \bgq \ values of the 
higher-redshift LBQS quasars.

\section{DISCUSSION}
\label{discussion}
In this section, we discuss our analysis and results in the 
context of previous work.
We can compare our analysis of the \Fb \ Archive data with that of 
\citet{fisher96} because we utilize the exact same data for 
19 of the 20 quasars they imaged.  
Our radio-loud subsamples are identical, while Fisher et al. include one more 
quasar (HE~1029$-$1401 at $z=0.086$) in their radio-quiet subsample.  
Fisher et 
al. do not list any individual statistics; we know only that the average value 
of \bgq \ for their
entire sample, considering only $m_* -1 < m < m_* +2$ galaxies, is 
$75_{-15}^{+18}~(h^{-1}_{100} \ \rm Mpc)^{1.77}$.  
For the same magnitude slice, our value
is $60 \pm 10~(h^{-1}_{100} \ \rm Mpc)^{1.77}$.  
For the radio-loud and radio-quiet subsamples, their average values 
of \bgq \ are $84_{-27}^{+33}$ and $72_{-19}^{+20}~(h^{-1}_{100} \ 
\rm Mpc)^{1.77}$, respectively.  In comparison,
our values are $66 \pm 18$ and $57 \pm 12~(h^{-1}_{100} \ \rm Mpc)^{1.77}$.
Although our results agree within the errors, our \bgq \ values are
systematically lower than the Fisher et al. values.  
The discrepancies translate into a difference of $\sim$1 
galaxy in the calculated average excess.  
Differences in how the expected
counts are estimated can probably account for the differences in \bgq \ values.  
For example, as mentioned in Section
\ref{num_counts}, the background counts estimated from the CNOC2
luminosity function may be systematically high by $< 0.5$ galaxy
per field for the F606W filter.  In addition, 
Fisher et al. do not use the PC data 
in their analysis, and they estimate the expected counts in WF3 using the
observed counts in WF2 and WF4.  

Another useful check of our analysis is to compare individual \bgq \ values with
those published by other authors.  
\citet{md00} also analyze the \Fb \ sample, and four of those
quasars are common to \citet{ye93}.  
Table \ref{indivbgq} compares \bgq \ for
these four quasars plus one additional \Fc \ quasar that is not in the
McLure \& Dunlop sample but is in the Yee \& Ellingson sample.
The agreement between our values and those
of \citet{ye93} is surprisingly good, considering the following
differences: they
use ground-based data covering a much wider field than WFPC2; they
estimate field galaxy counts from control fields; and they use different
parameters for calculating the luminosity function at the quasar redshift.  
\citet{md00} have analyzed the same 
$HST$ data as we, so one might expect better agreement between our
\bgq \ values.  However, \citet{md00} consider only galaxies
located on the same chip as the quasar (WF3) and estimate expected counts based
on the number of galaxies on WF2 and WF4.  Furthermore, they use different 
parameters when calculating the luminosity function at the quasar redshift.  
This comparison underscores how sensitive individual \bgq \ values are to the
methodology and indicates that results should only be 
interpreted in a statistical sense.

Finding reasonable agreement among our analysis and those
of \citet{fisher96}, \citet{md00}, and \citet{ye93}, 
we now discuss the significance of
our findings in the context of previous research. 
Our two main results are that 1) the LBQS
quasars lie in less dense environments than the more luminous Archive quasars,
and 2) radio-loud 
and radio-quiet quasars are found in similar environments. 
Table \ref{rlrqbgq} compares our average values of \bgq \  to those of other 
studies.  Note that the Archive \Fb \ and \citet{fisher96} data and samples
are identical except for one quasar, and \citet{md00} 
include all the \Fb \ quasars in their sample.
With this in mind, comparing \bgq \ values puts the findings of this 
study in a different light.  
First, the relatively sparse environments of the radio-quiet
LBQS quasars are consistent with previous ground-based studies,
but the radio-loud LBQS quasars are in unusually sparse environments
when compared to other radio-loud quasars.
Second, the richer environments of the Archive radio-loud
quasars are consistent with previous ground-based studies,
but the Archive radio-quiet quasars
are in unusually dense environments compared with 
the radio-quiet quasars studied by Smith et al. (1995) 
and \citet{eyg91}.
Furthermore, by comparing our values of \bgq \ to the \citet{ls79} 
value of 90~(\h~Mpc)$^{1.77}$ for Abell 0 clusters, 
we find that the Archive quasars
are located in galaxy environments slightly less rich than Abell 0 clusters.
When compared with the galaxy-galaxy covariance amplitude
of $20$~(\h~Mpc)$^{1.77}$ from \citet{dp83},
the LBQS quasars are in environments 
comparable to the typical galaxy.
This is not surprising for the radio-quiet LBQS subsample, but
one might expect the radio-loud LBQS quasars to be
in denser environments.

\begin{figure}
\caption{$B_{gq}$ versus $M_V$ for the entire sample of quasars.
Radio-loud quasars are represented with filled symbols,
and radio-quiet quasars are represented with open symbols.  
The LBQS quasars are represented
by triangles, and the $F606W$ and $F702W$ quasars are 
represented by squares and pentagons,
respectively.  There is no correlation between density of environment and $M_V$.}
\label{magbgq}
\epsscale{1.0}
\plotone{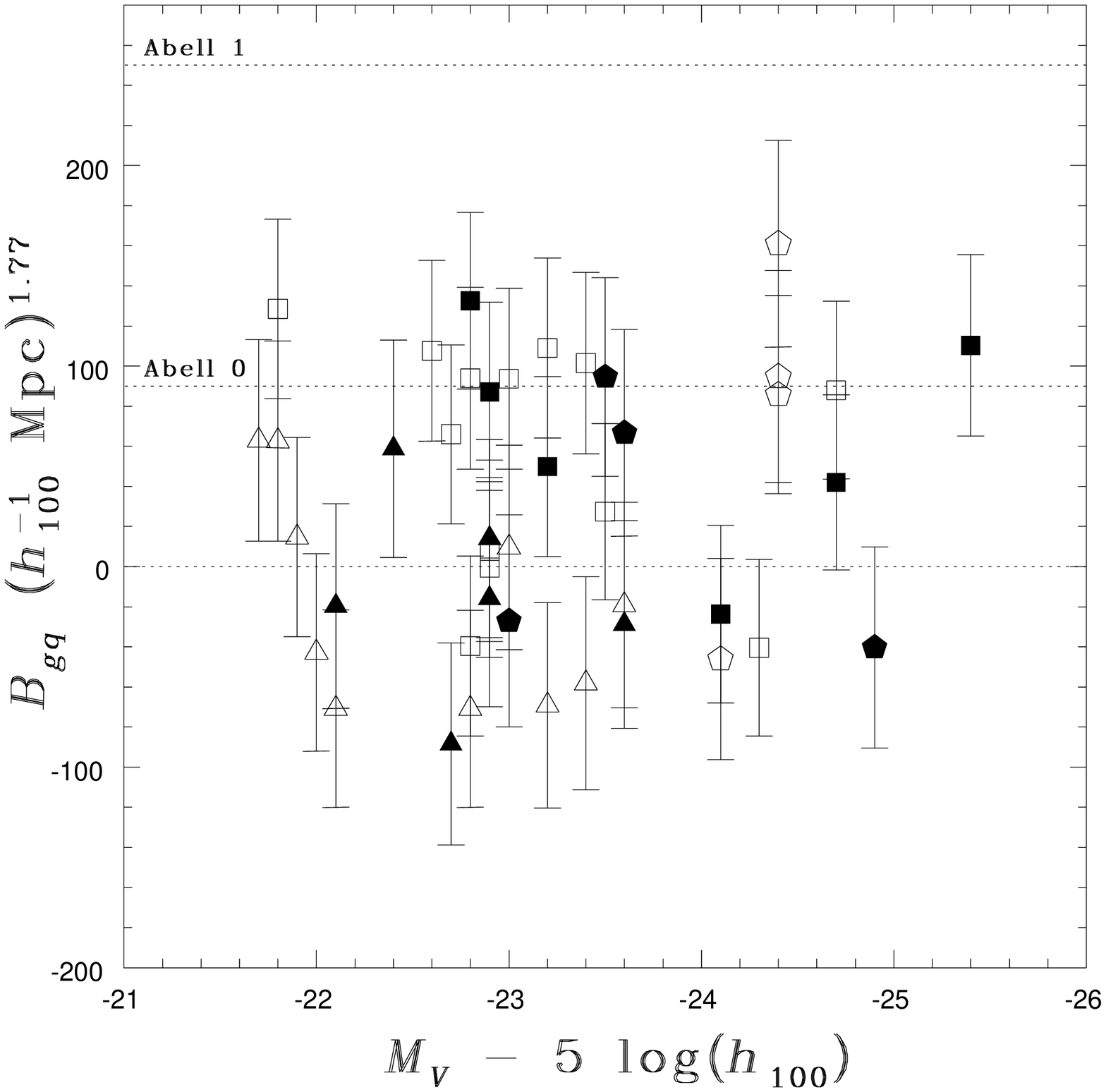}
\end{figure}

How do we make sense of these results?
\citet{fisher96} note that
the \Fb \ quasars are among the most luminous, and this might explain
why the \Fb \ radio-quiet quasars have denser environments
than average.  Furthermore,
the LBQS quasars are less luminous on average than the Archive quasars,
so maybe this explains why the LBQS are in relatively sparse environments.
Figure \ref{magbgq} compares \bgq \ with $M_V$, showing clearly
that environment is not correlated with optical luminosity.  
A Spearman rank test confirms this with a 60\% probability that
\bgq \ and $M_V$ are uncorrelated.
\citet{eyg91} look for a correlation between 
optical luminosity and environment with a sample of 96 quasars,
and they also find no correlation.  Therefore, it appears that
optical luminosity can not explain why the Archive radio-quiet
are in dense environments nor why the LBQS radio-loud quasars are in sparse 
environments. 
Radio luminosity might help explain the sparse environments of the
radio-loud LBQS quasars because
the radio-loud LBQS quasars have lower radio luminosities than the Archive 
radio-loud quasars.  To test this we
compare radio power at 5~GHz with \bgq.  Radio power is listed in 
Tables 1 and 2 for the LBQS and Archive quasars, respectively.
The LBQS 8.4~GHz
data from \citet{hif97} are converted to 5~GHz using a
spectral index of $-$0.32. Radio data for the \Fb \ sample are
from \citet{md00}, and data for the \Fc \ sample are from 
NASA/IPAC Extragalactic Database. 

\begin{figure}
\caption{Radio power at 5GHz versus $B_{gq}$ for quasars
with available radio data.  
The LBQS quasars are represented
by triangles and upper limits, and the $F606W$ and $F702W$ quasars are 
represented by squares and pentagons, respectively.
We find no
correlation between radio power and environment when considering
both radio-loud and quiet quasars.  However, when considering
the radio-loud quasars only, a Spearman rank test indicates
a 95\% probability that radio power is positively correlated with environment.}
\label{radio}
\epsscale{1.0}
\plotone{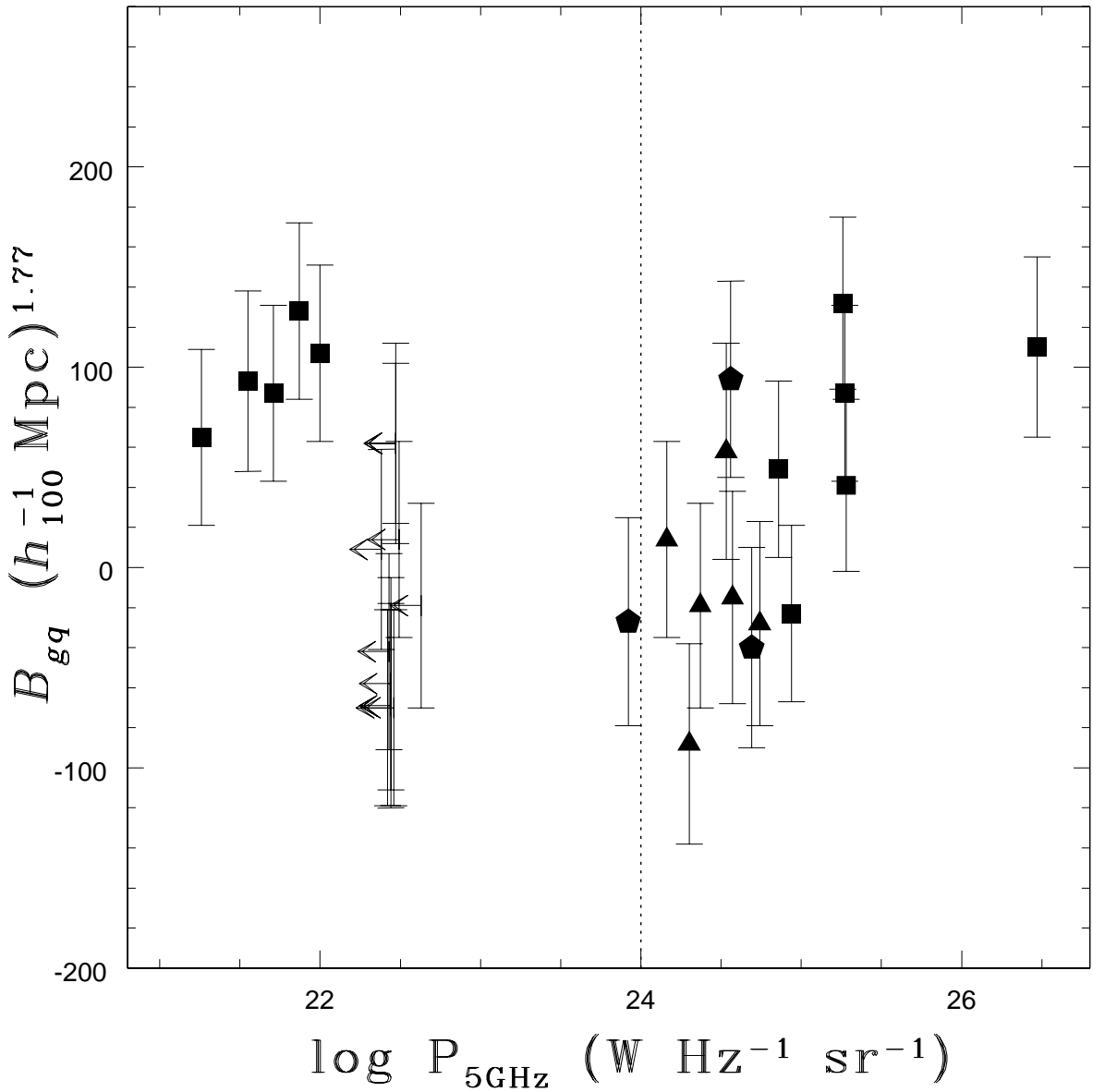}
\end{figure}

Figure \ref{radio} shows \bgq \ versus radio power, and
we find no
correlation between radio power and environment when considering
both radio-loud and quiet quasars.  We note that
almost all the LBQS quasars are a factor of
10-100 less powerful that the radio-loud
Archive quasars, and it is not clear that
the same emission mechanism holds across this
large range in radio power.  However, when
considering the radio-loud quasars only, a Spearman rank test indicates a
modest correlation (95\% probability) between radio power and environment,
which is dependent on the one point at extreme values of radio luminosity
and \bgq \ (85\% probability of a correlation if this point is removed).
\citet{wold00} find a slight but significant correlation between
radio power and environment, although they consider steep
spectrum sources only.  Most of the radio-loud quasars in our samples
are flat spectrum, yet the correlation remains.  Interestingly,
\citet{md00} note that a weak trend between radio power
and environment is suggested by their data, but the correlation
is not statistically significant.
The richest environments in the \citet{yg87} and \citet{eyg91} samples
are around very radio luminous, steep-spectrum sources.  The addition
of their data to Figure \ref{radio} might strengthen the correlation
between environment and radio power, but we resist adding their numbers
due to possible systematic differences in analysis.
Since most of the radio data for the radio-quiet quasars are upper limits,
we cannot test for an independent correlation between
radio power and environment for radio-quiet quasars.  
The fact that the correlation in the radio-loud data does not 
apply to both radio-loud and quiet 
quasars implies that two emission mechanisms may be at
work \citep[e.g.][]{stocke92, hooper96}, even though there 
appears to be a continuum of radio power among quasars.

Finally, what is the true incidence of clustering around quasars?
As discussed in Section \ref{introduction}, the LBQS is 
among the most representative surveys of the currently known
radio-quiet quasar population, and the radio-quiet quasars
make up $\approx$90\% of the whole.
Therefore, most quasars, like the LBQS sample
presented here, lie in environments
comparable to the typical galaxy.
The Archive
radio-quiet quasars have unusually dense environments and are thus
not a representative sample of radio-quiet quasars.  The clustering
associated with radio-loud quasars correlates with radio power, and 
the environments of the radio-loud quasars presented here 
range from that of a typical galaxy to Abell 0 clusters.   
We present data for only 16 LBQS quasars, and a larger sample is 
needed to strengthen these results.
In addition, spectroscopic studies provide more insight 
into the dynamics of quasar environments, enabling
one to examine variations in 
environments as opposed to limiting analysis to the average properties
of the sample.  The fluctuations in field galaxy counts preclude the analysis
of individual quasars using our current method.

The LBQS makes an excellent sample for extending quasar environment
studies to higher redshift to look for evolution.
This work has been started by \citet{wold01}, who include
10 LBQS radio-quiet quasars in their study of $0.5 \le z \le 0.8$ 
radio-quiet quasar environments.  \citet{wold01} conclude that,
on average, $0.5 \le z \le 0.8$ radio-quiet 
are found in environments that are $3 \times$ more 
dense than the typical galaxy.  
However, the 10 LBQS quasars in their radio-quiet
sample have environments comparable to the typical galaxy (for their
star-subtraction Models 2 and 3).
We have ground-based $R$ and $H$-band data for a larger sample of LBQS quasars,
which includes the sample presented here as well as higher redshift quasars.
The ground-based data are much deeper than the \hst \ data, 
and their analysis will 
help elucidate the true incidence of clustering around quasars
as a function of redshift.

\section{SUMMARY}
\label{conclusions}
We present $Hubble~Space~Telescope$ ($HST$) Wide Field Planetary Camera 2 (WFPC2) 
data on the large-scale environments of
16 $0.39 < z < 0.51$ quasars from the Large Bright Quasar Survey (LBQS).  
This is the first look at the large-scale environments of LBQS quasars,
and this is significant because the LBQS quasars are representative of the radio-quiet 
quasar population.  
We compare the LBQS environments with the environments of 27 
$0.15 < z < 0.55$ quasars selected from the $HST$ Archive.
The analysis of the Archive sample is useful for two reasons.
First, the Archive sample provides a check of our methodology because
most of these data have been published in previous environment studies.  
Second, the majority of the Archive quasars are 
from the PG and PKS surveys, and these quasars are more luminous on 
average than the LBQS quasars; by comparing the LBQS and Archive
studies we investigate whether previous quasar environment studies have
been biased due to studying unusually radio or optically luminous quasars.
To quantify the quasar environments, we compare observed galaxy number counts with 
expected counts predicted from the CNOC2 field-galaxy luminosity function
in order to look for statistical excesses of galaxies around the quasars.  
We detect a significant excess of galaxies around the Archive quasars 
but find no such excess around the LBQS quasars.
We calculate the amplitude of the spatial correlation function, and 
we find that the LBQS environments are consistent with that of the typical
galaxy while the Archive environments are
slightly less rich than Abell 0 clusters. 
Contrary to previous ground-based studies,
we find no difference between the environments of radio-loud and radio-quiet
quasars of either sample.  However, comparison of \bgq \ values with previously
published values shows the radio-loud LBQS quasars are in environments
less dense than most other radio-loud quasars.  In addition, the 
Archive radio-quiet quasars are in anomalously dense environments compared
to other radio-quiet quasars. 
The richer environments of the Archive radio-quiet quasars can not
be explained by their higher optical luminosities because we find
no correlation between optical luminosity and environment.  
We do find a positive correlation (95\%) between radio luminosity and environment
for the radio-loud quasars.  The LBQS radio-loud quasars have lower radio
powers than the radio-loud Archive quasars, and this might explain why the LBQS radio-loud
quasars are in sparser environments.

\acknowledgements
The authors would like to thank H. Lin for kind assistance in 
utilizing CNOC2 results and for comparing PPP with SExtractor.  
This work was partially supported by NASA through GO program 5450 from 
the Space Telescope Science Institute, which is operated by 
the Association of Universities 
for Research in Astronomy, Inc., under NASA contract NAS 5-26555.
This research has made use of
the NASA/IPAC Extragalactic Database (NED) which is operated by 
the Jet Propulsion Laboratory, California Institute of Technology, under
contract with the National Aeronautics and Space Administration. 
RAF acknowledges support from the UA/NASA Spacegrant Fellowship, NSF
grant AST-9623788, and NGT-5-50283, the latter a NASA Graduate
Student Researchers Program Fellowship.
EJH acknowledges support from NASA grants NAGW-3134 and NGT-51152, 
the latter a NASA Graduate Student Researchers Program Fellowship 
at the University of Arizona.

\newpage

\begin{deluxetable}{lrrrcrccc} 
\tablenum{1} 
\tablewidth{0pt} 
\tablecaption{LBQS Sample} 
\tablehead{\colhead{Name} & \colhead{z} & \colhead{$B_J$\tablenotemark{a}} & \colhead{$M_{V}$\tablenotemark{b}} & \colhead{RL/RQ} & \colhead{log(P$\rm_{5GHz}$\tablenotemark{c}~)} & \colhead{Filter} & \colhead{t$\rm _{total}$} & \colhead{N$\rm_{exp}$} \\
 & & & & & & & \colhead{(sec)} &  }  
\startdata  
{0020+0018}  & 0.4232 & 18.60  & $-$21.83 & RQ  &$<$22.47 & F675W & 1200     & 3  \\ 
{0021$-$0301}& 0.4218 & 18.77  & $-$21.65 & RQ  &$<$22.47 & 	 & 1200     & 3  \\ 
{0100+0205}  & 0.3934 & 17.51 &  $-$22.78 & RQ  &$<$22.42 & 	 & 1280     & 4   \\
{1138+0003}  & 0.5003 & 17.90 &  $-$22.85 & RL  &   24.57    & 	 & 1280     & 4   \\
{1149+0043}  & 0.4661 & 17.04 &  $-$23.56 & RQ  &$<$22.63 & 	 & 1480     & 5   \\
{1209+1259}  & 0.4180 & 18.55 &  $-$21.86 & RQ  &$<$22.49 & 	 & 1200     & 3   \\ 
{1218+1734}  & 0.4445 & 17.78 &  $-$22.74 & RL  &   24.30    & 	 & 1200     & 3   \\
{1222+1010}  & 0.3978 & 18.33 &  $-$21.97 & RQ  &$<$22.43 & 	 & 1200     & 3   \\ 
{1222+1235}  & 0.4121 & 17.43 &  $-$22.94 & RL  & 24.16      & 	 & 1200     & 3   \\
{1230$-$0015}& 0.4705 & 17.00 &  $-$23.62 & RL  & 24.64      & 	 & 1580     & 5   \\
{1240+1754}  & 0.4584 & 17.41 &  $-$23.16 & RQ  &$<$22.44 & 	 & 1480     & 5   \\
{1242$-$0123}& 0.4906 & 17.33 &  $-$23.38 & RQ  &$<$22.44 & 	 & 1280     & 4   \\
{1243+1701}  & 0.4591 & 17.61 &  $-$22.97 & RQ  &$<$22.38 & 	 & 1400     & 4   \\ 
{2214$-$1903}& 0.3965 & 18.16 &  $-$22.14 & RQ  &$<$22.46 & 	 & 1280     & 4   \\ 
{2348+0210}  & 0.5039 & 18.35 &  $-$22.41 & RL  &   24.53    & 	 & 1280     & 4   \\
{2351$-$0036}& 0.4600 & 18.47 &  $-$22.11 & RL  &   24.37    & 	 & 1200     & 3   \\
\enddata
\tablenotetext{a}{$B_J$ magnitudes are from \citet{hfc95}.}
\tablenotetext{b}{$M_V$ is calculated from $B_J$ using color and k-corrections from \citet{cv90}.}
\tablenotetext{c}{Units of $\rm W~Hz^{-1}~sr^{-1}$.  Assumes $H_0 = 50$~km~s$^{-1}$~Mpc$^{-1}$.  The 8.4~GHz data from \citet{hif97} are converted to 5~GHz using a spectral index of $-$0.32.}
\end{deluxetable}

\begin{deluxetable}{lrrrcrcccc} 
\tablenum{2} 
\tablewidth{0pt} 
\tablecaption{HST Archive Sample} 
\tablehead{\colhead{Name} & \colhead{z} & \colhead{$V$\tablenotemark{a}} & \colhead{$M_{V}$\tablenotemark{b}} & \colhead{RL/RQ} & \colhead{log(P$\rm_{5GHz}$\tablenotemark{c}~)}& \colhead{Filter} & \colhead{t$\rm _{total}$} & \colhead{N$\rm_{exp}$} & \colhead{Ref.} \\
 & & & & & & & \colhead{(sec)} & &}  
\startdata  
PG~0052+251 	& 0.155  & 15.42 &  $-$22.8     	& RQ & 21.55	& F606W & 2100 & 3 & 1 \\ 
HB89~0205+024 	& 0.155  & 15.39 &  $-$22.9  		& RQ & \nodata  &    & 2100 & 3    &    \\ 
Q~0316$-$346 	& 0.265  & 15.2\phn  &  $-$24.3 	& RQ & \nodata 	&& 2100 & 3 	&   \\ 
PG~0923+201 	& 0.190  & 16.04 &  $-$22.7  		& RQ & 21.26	&& 2100 & 3 	&    \\ 
PG~0953+414 	& 0.234  & 14.5\phn  &  $-$24.7 	& RQ & 21.71 	&& 1800 & 3 &  \\ 
PKS~1004+13 	& 0.240  & 15.15 &  $-$24.1  		& RL & 24.94 	&& 2100 & 3 &  \\ 
PG~1012+008 	& 0.185  & 16.0\phn  &  $-$22.6 	& RQ & 22.00 	&& 2100 & 3 &  \\ 
PG~1116+215 	& 0.177  & 15.17 &  $-$23.4  		& RQ & \nodata 	&& 1800 & 3 &  \\ 
PG~1202+281 	& 0.165  & 15.51 &  $-$22.9  		& RQ & \nodata 	&& 1800 & 3 &  \\ 
PG~1226+023\tablenotemark{d} & 0.158  & 12.86 &  $-$25.4& RL & 26.47 	&& 1800 & 3 &  \\ 
PKS~1302$-$102 	& 0.286  & 14.92 &  $-$24.7  		& RL & 25.28 	&& 1800 & 3 &  \\ 
PG~1307+085 	& 0.155  & 15.28 &  $-$23.0  		& RQ & \nodata 	&& 2100 & 3 &  \\ 
PG~1309+355 	& 0.184  & 15.45 &  $-$23.2  		& RQ & \nodata 	&& 2100 & 3 &  \\ 
PG~1402+261 	& 0.164  & 15.57 &  $-$22.8  		& RQ & \nodata 	&& 2100 & 3 &  \\ 
PG~1444+407 	& 0.267  & 15.95 &  $-$23.5  		& RQ & \nodata 	&& 1800 & 3 &  \\ 
PG~1545+210\tablenotemark{e}	& 0.266  & 16.69 &  $-$22.8  & RL & 25.26& 	& 1800 & 3 &    \\ 
PKS~2135$-$14 	& 0.200  & 15.91 &  $-$22.9  		& RL & 25.27 	&& 2100 & 3 &    \\ 
PKS~2349$-$01 	& 0.173  & 15.33 &  $-$23.2  		& RL & 24.86 	&& 2100 & 3 &    \\ 
PHL~909 	& 0.171  & 16.7\phn  &  $-$21.8  	& RQ & 21.87 	&& 2100 & 3 &    \\ 
\\				    	    		  	      
PG~0043+039 	& 0.385  & 15.88 &  $-$24.4 		& RQ &\nodata &F702W	& 1800 & 4 & 2  \\ 
PKS~0202$-$76 	& 0.389  & 16.77 &  $-$23.5 		& RL &24.56 & 	& 1800 & 4         & 2  \\ 
PKS~0312$-$77 	& 0.223  & 16.10 &  $-$23.0 		& RL &23.92 & 	& 1800 & 4         & 2  \\ 
IRAS~04505$-$2958 & 0.286 & 16.0\phn  &  $-$23.6 	& RL &\nodata & 	& 1800 & 4 & 3   \\ 
HB~0850+440 	& 0.514  & 16.4\phn  &  $-$24.4 	& RQ &\nodata & 	& 2400 & 3 & 4 \\ 
PG~1001+291 	& 0.330  & 15.51 &  $-$24.4 		& RQ &\nodata & 	& 2400 & 3 & 5 \\ 
PG~1358+043 	& 0.427  & 16.31 &  $-$24.1 		& RQ &\nodata & 	& 1800 & 4 & 2  \\ 
PG~1704+608 	& 0.372  & 15.28 &  $-$24.9 		& RL &24.69 & 	& 1800 & 4         & 2  \\ 
\enddata  
\tablenotetext{a}{$V$ magnitudes are from \citet{hb89}.}
\tablenotetext{b}{$M_V$ is calculated from $V$ using k-corrections from \citet{cv90}.}
\tablenotetext{c}{Units of $\rm W~Hz^{-1}~sr^{-1}$.  Assumes $H_0 = 50$~km~s$^{-1}$~Mpc$^{-1}$.  
The \Fb \ data are from \citet{md00}.}

\tablenotetext{d}{3C273}
\tablenotetext{e}{3C323.1}
\tablerefs{(1) Fisher et al. 1996; (2) Boyce et al. 1998; (3) Disney et al. 1995; (4) Lanzetta et al. 1997; (5) Boyce, Disney, \& Bleaden 1999} 
\end{deluxetable} 

\begin{deluxetable}{lcrrrrrrrrrrrrr} 
\rotate 
\tablenum{3} 
\tabletypesize{\tiny} 
\tablewidth{0pt} 
\tablecolumns{13} 
\tablecaption{Galaxy Counts and Clustering for LBQS Sample \label{results_lbqs}} 
\tablehead{\colhead{Name} & \colhead{z} & 
\colhead{$\theta_{max}$} & \colhead{r$_{max}$} 
& \multicolumn{5}{c}{$14 < m < 22$} & 
\multicolumn{6}{c}{$m_* - 1 < m < m_* +2$}  \\ 
\colhead{$$}    & \colhead{$$}   & \colhead{(\arcsec)} 
& \colhead{($h^{-1}_{100} \ \rm kpc $)} & \colhead{N$_{obs}$} 
& \colhead{N$_{exp}$} & \colhead{N$_{excess}$} & \colhead{$\delta$} 
& \colhead{B$_{gq}$\tablenotemark{a}} & \colhead{m$_*$} & \colhead{N$_{obs}$} 
& \colhead{N$_{exp}$} & \colhead{N$_{excess}$} & \colhead{$\delta$} 
& \colhead{B$_{gq}$\tablenotemark{a}} }  
\startdata  
LBQS~0020+0018 & 0.423 & 130.3 &   430 &  25 & 17.8 $\pm$ 5.5 & 7.2 $\pm$ 5.5 &  0.40 &   65 $\pm$   49 & 20.4 &  23 & 16.4 $\pm$ 5.3 & 6.6 $\pm$ 5.3 &  0.40 &   62 $\pm$   49  \\ 
LBQS~0021$-$0301 & 0.422 & 131.7 &   434 &  25 & 17.8 $\pm$ 5.5 & 7.2 $\pm$ 5.5 &  0.40 &   65 $\pm$   50 & 20.4 &  23 & 16.4 $\pm$ 5.3 & 6.6 $\pm$ 5.3 &  0.40 &   62 $\pm$   50  \\ 
LBQS~0100+0205 & 0.393 & 130.6 &   416 &  10 & 17.8 $\pm$ 5.5 & -7.8 $\pm$ 5.5 & -0.44 &  -69 $\pm$   48 & 20.3 &   9 & 16.6 $\pm$ 5.3 & -7.6 $\pm$ 5.3 & -0.46 &  -70 $\pm$   49  \\ 
LBQS~1138+0003$^b$ & 0.500 & 130.4 &   464 &  16 & 17.8 $\pm$ 5.5 & -1.8 $\pm$ 5.5 & -0.10 &  -18 $\pm$   55 & 20.9 &  14 & 15.5 $\pm$ 5.1 & -1.5 $\pm$ 5.1 & -0.10 &  -15 $\pm$   53  \\ 
LBQS~1149+0043 & 0.466 & 128.9 &   445 &  14 & 17.8 $\pm$ 5.5 & -3.8 $\pm$ 5.5 & -0.22 &  -36 $\pm$   51 & 20.7 &  14 & 15.9 $\pm$ 5.2 & -1.9 $\pm$ 5.2 & -0.12 &  -19 $\pm$   51  \\ 
LBQS~1209+1259 & 0.418 & 129.9 &   426 &  22 & 17.8 $\pm$ 5.5 & 4.2 $\pm$ 5.5 &  0.23 &   37 $\pm$   49 & 20.4 &  18 & 16.4 $\pm$ 5.3 & 1.6 $\pm$ 5.3 &  0.10 &   14 $\pm$   49  \\ 
LBQS~1218+1734$^b$ & 0.444 & 129.8 &   438 &   8 & 17.8 $\pm$ 5.5 & -9.8 $\pm$ 5.5 & -0.55 &  -90 $\pm$   50 & 20.6 &   7 & 16.2 $\pm$ 5.2 & -9.2 $\pm$ 5.2 & -0.57 &  -88 $\pm$   50  \\ 
LBQS~1222+1010 & 0.398 & 130.3 &   418 &  12 & 17.8 $\pm$ 5.5 & -5.8 $\pm$ 5.5 & -0.33 &  -52 $\pm$   49 & 20.3 &  12 & 16.6 $\pm$ 5.3 & -4.6 $\pm$ 5.3 & -0.28 &  -42 $\pm$   49  \\ 
LBQS~1222+1235$^b$ & 0.412 & 130.1 &   424 &  20 & 17.8 $\pm$ 5.5 & 2.2 $\pm$ 5.5 &  0.12 &   19 $\pm$   49 & 20.4 &  18 & 16.5 $\pm$ 5.3 & 1.5 $\pm$ 5.3 &  0.09 &   14 $\pm$   49  \\ 
LBQS~1230$-$0015$^b$ & 0.470 & 130.1 &   451 &  14 & 17.8 $\pm$ 5.5 & -3.8 $\pm$ 5.5 & -0.22 &  -36 $\pm$   52 & 20.7 &  13 & 15.9 $\pm$ 5.2 & -2.9 $\pm$ 5.2 & -0.18 &  -28 $\pm$   51  \\ 
LBQS~1240+1754 & 0.458 & 130.3 &   447 &   9 & 17.8 $\pm$ 5.5 & -8.8 $\pm$ 5.5 & -0.50 &  -83 $\pm$   51 & 20.6 &   9 & 16.0 $\pm$ 5.2 & -7.0 $\pm$ 5.2 & -0.44 &  -69 $\pm$   51  \\ 
LBQS~1242$-$0123 & 0.491 & 130.2 &   459 &  14 & 17.8 $\pm$ 5.5 & -3.8 $\pm$ 5.5 & -0.22 &  -37 $\pm$   54 & 20.8 &  10 & 15.6 $\pm$ 5.1 & -5.6 $\pm$ 5.1 & -0.36 &  -58 $\pm$   53  \\ 
LBQS~1243+1701 & 0.459 & 129.1 &   443 &  18 & 17.8 $\pm$ 5.5 & 0.2 $\pm$ 5.5 &  0.01 &    1 $\pm$   51 & 20.6 &  17 & 16.0 $\pm$ 5.2 & 1.0 $\pm$ 5.2 &  0.06 &    9 $\pm$   50  \\ 
LBQS~2214$-$1903 & 0.396 & 130.4 &   417 &  10 & 17.8 $\pm$ 5.5 & -7.8 $\pm$ 5.5 & -0.44 &  -70 $\pm$   49 & 20.3 &   9 & 16.6 $\pm$ 5.3 & -7.6 $\pm$ 5.3 & -0.46 &  -70 $\pm$   49  \\ 
LBQS~2348+0210$^b$ & 0.504 & 130.6 &   466 &  24 & 17.8 $\pm$ 5.5 & 6.2 $\pm$ 5.5 &  0.35 &   62 $\pm$   55 & 20.9 &  21 & 15.5 $\pm$ 5.1 & 5.5 $\pm$ 5.1 &  0.36 &   58 $\pm$   54  \\ 
LBQS~2351$-$0036$^b$ & 0.460 & 129.4 &   444 &  16 & 17.8 $\pm$ 5.5 & -1.8 $\pm$ 5.5 & -0.10 &  -17 $\pm$   51 & 20.7 &  14 & 16.0 $\pm$ 5.2 & -2.0 $\pm$ 5.2 & -0.13 &  -19 $\pm$   51  \\ 
\hline  
Average &  &  &  & 16.1 $\pm$ 5.6 & & -1.8 $\pm$ 1.4\tablenotemark{c} & -0.10 &  -16 $\pm$   12\tablenotemark{c} & & 14.4 $\pm$ 5.0 & & -1.7 $\pm$ 1.3\tablenotemark{c} & -0.10 &  -16 $\pm$   12\tablenotemark{c}  \\ 
Radio Loud &  &  & & & &  & &  -13 $\pm$   21\tablenotemark{c} &  & &  &  & &  -13 $\pm$   21\tablenotemark{c}  \\ 
Radio Quiet &  &  & & & & &  &  -17 $\pm$   16\tablenotemark{c} &  & &  & &  &  -18 $\pm$   15\tablenotemark{c}  \\ 
\enddata  
\tablenotetext{a}{Units of $(h^{-1}_{100} \ \rm Mpc)^{1.77}$. Error calculations are described in text.} 
\tablenotetext{b}{Radio-Loud Quasars} 
\tablenotetext{c}{Error is $\frac{\sqrt{\sum \sigma_i^2}}{n} $} 
\end{deluxetable}

\begin{deluxetable}{lcrrrrrrrrrrrrr} 
\rotate 
\tablenum{4} 
\tabletypesize{\tiny} 
\tablewidth{0pt} 
\tablecolumns{13} 
\tablecaption{Galaxy Counts and Clustering for $F606W$ Archive Sample \label{results_f606w}} 
\tablehead{\colhead{Name} & \colhead{z} & 
\colhead{$\theta_{max}$} & \colhead{r$_{max}$} 
& \multicolumn{5}{c}{$14 < m < 22$} & 
\multicolumn{6}{c}{$m_* - 1 < m < m_* +2$}  \\ 
\colhead{$$}    & \colhead{$$}   & \colhead{(\arcsec)} 
& \colhead{($h^{-1}_{100} \ \rm kpc $)} & \colhead{N$_{obs}$} 
& \colhead{N$_{exp}$} & \colhead{N$_{excess}$} & \colhead{$\delta$} 
& \colhead{B$_{gq}$\tablenotemark{a}} & \colhead{m$_*$} & \colhead{N$_{obs}$} 
& \colhead{N$_{exp}$} & \colhead{N$_{excess}$} & \colhead{$\delta$} 
& \colhead{B$_{gq}$\tablenotemark{a}} }  
\startdata  
PG~0052+251 & 0.155 & 119.0 &   208 &  19 & 10.9 $\pm$ 4.3 & 8.1 $\pm$ 4.3 &  0.74 &   93 $\pm$   49 & 18.5 &   7 & 2.6 $\pm$ 2.1 & 4.4 $\pm$ 2.1 &  1.66 &   93 $\pm$   45  \\ 
HB89~0205+024 & 0.155 & 119.6 &   209 &  10 & 10.9 $\pm$ 4.3 & -0.9 $\pm$ 4.3 & -0.08 &  -10 $\pm$   49 & 18.5 &   3 & 2.6 $\pm$ 2.1 & 0.4 $\pm$ 2.1 &  0.14 &    7 $\pm$   45  \\ 
Q~0316$-$346 & 0.265 & 119.6 &   304 &   6 & 10.9 $\pm$ 4.3 & -4.9 $\pm$ 4.3 & -0.45 &  -49 $\pm$   43 & 19.8 &   5 & 8.5 $\pm$ 3.8 & -3.5 $\pm$ 3.8 & -0.41 &  -40 $\pm$   44  \\ 
PG~0923+201 & 0.190 & 118.1 &   240 &  20 & 10.9 $\pm$ 4.3 & 9.1 $\pm$ 4.3 &  0.83 &   96 $\pm$   45 & 19.0 &   8 & 4.1 $\pm$ 2.6 & 3.9 $\pm$ 2.6 &  0.95 &   65 $\pm$   44  \\ 
PG~0953+414 & 0.234 & 121.3 &   285 &  23 & 10.9 $\pm$ 4.3 & 12.1 $\pm$ 4.3 &  1.11 &  122 $\pm$   43 & 19.5 &  13 & 6.5 $\pm$ 3.3 & 6.5 $\pm$ 3.3 &  1.02 &   87 $\pm$   44  \\ 
PKS~1004+13$^b$ & 0.240 & 119.6 &   286 &   5 & 10.9 $\pm$ 4.3 & -5.9 $\pm$ 4.3 & -0.54 &  -60 $\pm$   43 & 19.5 &   5 & 6.8 $\pm$ 3.4 & -1.8 $\pm$ 3.4 & -0.27 &  -23 $\pm$   44  \\ 
PG~1012+008 & 0.185 & 119.4 &   238 &  21 & 10.9 $\pm$ 4.3 & 10.1 $\pm$ 4.3 &  0.92 &  108 $\pm$   46 & 18.9 &  10 & 3.9 $\pm$ 2.6 & 6.1 $\pm$ 2.6 &  1.58 &  107 $\pm$   44  \\ 
PG~1116+215 & 0.177 & 121.9 &   235 &  20 & 10.9 $\pm$ 4.3 & 9.1 $\pm$ 4.3 &  0.83 &   99 $\pm$   47 & 18.8 &   9 & 3.5 $\pm$ 2.4 & 5.5 $\pm$ 2.4 &  1.56 &  101 $\pm$   45  \\ 
PG~1202+281 & 0.165 & 121.8 &   223 &  17 & 10.9 $\pm$ 4.3 & 6.1 $\pm$ 4.3 &  0.56 &   67 $\pm$   47 & 18.7 &   3 & 3.0 $\pm$ 2.3 & -0.0 $\pm$ 2.3 & -0.01 &    0 $\pm$   44  \\ 
PG~1226+023$^b$ & 0.158 & 119.8 &   212 &  20 & 10.9 $\pm$ 4.3 & 9.1 $\pm$ 4.3 &  0.83 &  103 $\pm$   49 & 18.6 &   8 & 2.7 $\pm$ 2.2 & 5.3 $\pm$ 2.2 &  1.91 &  110 $\pm$   45  \\ 
PKS~1302$-$102$^b$ & 0.286 & 119.4 &   319 &  15 & 10.9 $\pm$ 4.3 & 4.1 $\pm$ 4.3 &  0.37 &   41 $\pm$   43 & 20.0 &  14 & 10.0 $\pm$ 4.1 & 4.0 $\pm$ 4.1 &  0.39 &   41 $\pm$   43  \\ 
PG~1307+085 & 0.155 & 123.1 &   215 &  24 & 10.9 $\pm$ 4.3 & 13.1 $\pm$ 4.3 &  1.20 &  150 $\pm$   49 & 18.5 &   7 & 2.6 $\pm$ 2.1 & 4.4 $\pm$ 2.1 &  1.66 &   93 $\pm$   45  \\ 
PG~1309+355 & 0.184 & 119.2 &   237 &  17 & 10.9 $\pm$ 4.3 & 6.1 $\pm$ 4.3 &  0.56 &   65 $\pm$   46 & 18.9 &  10 & 3.8 $\pm$ 2.5 & 6.2 $\pm$ 2.5 &  1.61 &  108 $\pm$   44  \\ 
PG~1402+261 & 0.164 & 118.7 &   216 &   8 & 10.9 $\pm$ 4.3 & -2.9 $\pm$ 4.3 & -0.27 &  -32 $\pm$   48 & 18.7 &   1 & 3.0 $\pm$ 2.2 & -2.0 $\pm$ 2.2 & -0.66 &  -39 $\pm$   44  \\ 
PG~1444+407 & 0.267 & 119.3 &   305 &  11 & 10.9 $\pm$ 4.3 & 0.1 $\pm$ 4.3 &  0.01 &    0 $\pm$   43 & 19.8 &  11 & 8.6 $\pm$ 3.8 & 2.4 $\pm$ 3.8 &  0.28 &   27 $\pm$   43  \\ 
PG~1545+210$^b$ & 0.266 & 118.3 &   302 &  22 & 10.9 $\pm$ 4.3 & 11.1 $\pm$ 4.3 &  1.01 &  111 $\pm$   43 & 19.8 &  20 & 8.5 $\pm$ 3.8 & 11.5 $\pm$ 3.8 &  1.34 &  132 $\pm$   43  \\ 
PKS~2135$-$14$^b$ & 0.200 & 119.9 &   253 &  16 & 10.9 $\pm$ 4.3 & 5.1 $\pm$ 4.3 &  0.47 &   53 $\pm$   45 & 19.1 &  10 & 4.6 $\pm$ 2.8 & 5.4 $\pm$ 2.8 &  1.18 &   87 $\pm$   44  \\ 
PKS~2349$-$01$^b$ & 0.173 & 117.8 &   224 &  16 & 10.9 $\pm$ 4.3 & 5.1 $\pm$ 4.3 &  0.47 &   55 $\pm$   47 & 18.8 &   6 & 3.3 $\pm$ 2.4 & 2.7 $\pm$ 2.4 &  0.79 &   49 $\pm$   44  \\ 
PHL~909 & 0.171 & 118.8 &   223 &  21 & 10.9 $\pm$ 4.3 & 10.1 $\pm$ 4.3 &  0.92 &  110 $\pm$   47 & 18.8 &  10 & 3.3 $\pm$ 2.3 & 6.7 $\pm$ 2.3 &  2.06 &  128 $\pm$   44  \\ 
\hline  
Average &  &  &  & 16.4 $\pm$ 5.6 & & 5.4 $\pm$ 1.0\tablenotemark{c} &  0.50 &   59 $\pm$   10\tablenotemark{c} & & 8.4 $\pm$ 4.3 & & 3.6 $\pm$ 0.7\tablenotemark{c} &  0.88 &   60 $\pm$   10\tablenotemark{c}  \\ 
Radio Loud &  &  & & & &  & &   50 $\pm$   18\tablenotemark{c} &  & &  &  & &   66 $\pm$   18\tablenotemark{c}  \\ 
Radio Quiet &  &  & & & & &  &   63 $\pm$   12\tablenotemark{c} &  & &  & &  &   57 $\pm$   12\tablenotemark{c}  \\ 
\enddata  
\tablenotetext{a}{Units of $(h^{-1}_{100} \ \rm Mpc)^{1.77}$. Error calculations are described in text.} 
\tablenotetext{b}{Radio-Loud Quasars} 
\tablenotetext{c}{Error is $\frac{\sqrt{\sum \sigma_i^2}}{n} $} 
\end{deluxetable}

\begin{deluxetable}{lcrrrrrrrrrrrrr} 
\rotate 
\tablenum{5} 
\tabletypesize{\tiny} 
\tablewidth{0pt} 
\tablecolumns{13} 
\tablecaption{Galaxy Counts and Clustering for $F702W$ Archive Sample \label{results_f702w}} 
\tablehead{\colhead{Name} & \colhead{z} & 
\colhead{$\theta_{max}$} & \colhead{r$_{max}$} 
& \multicolumn{5}{c}{$14 < m < 22$} & 
\multicolumn{6}{c}{$m_* - 1 < m < m_* +2$}  \\ 
\colhead{$$}    & \colhead{$$}   & \colhead{(\arcsec)} 
& \colhead{($h^{-1}_{100} \ \rm kpc $)} & \colhead{N$_{obs}$} 
& \colhead{N$_{exp}$} & \colhead{N$_{excess}$} & \colhead{$\delta$} 
& \colhead{B$_{gq}$\tablenotemark{a}} & \colhead{m$_*$} & \colhead{N$_{obs}$} 
& \colhead{N$_{exp}$} & \colhead{N$_{excess}$} & \colhead{$\delta$} 
& \colhead{B$_{gq}$\tablenotemark{a}} }  
\startdata  
PG~0043+039 & 0.385 & 130.2 &   410 &  29 & 19.5 $\pm$ 5.7 & 9.5 $\pm$ 5.7 &  0.49 &   81 $\pm$   49 & 20.1 &  28 & 18.4 $\pm$ 5.6 & 9.6 $\pm$ 5.6 &  0.52 &   85 $\pm$   49  \\ 
PKS~0202$-$76$^b$ & 0.389 & 129.2 &   409 &  30 & 19.5 $\pm$ 5.7 & 10.5 $\pm$ 5.7 &  0.54 &   89 $\pm$   49 & 20.2 &  29 & 18.4 $\pm$ 5.6 & 10.6 $\pm$ 5.6 &  0.58 &   94 $\pm$   49  \\ 
PKS~0312$-$77$^b$ & 0.223 & 130.2 &   296 &  10 & 19.5 $\pm$ 5.7 & -9.5 $\pm$ 5.7 & -0.49 &  -88 $\pm$   53 & 18.9 &   5 & 6.7 $\pm$ 3.4 & -1.7 $\pm$ 3.4 & -0.26 &  -27 $\pm$   52  \\ 
IRAS~04505$-$2958$^b$ & 0.286 & 129.6 &   346 &  34 & 19.5 $\pm$ 5.7 & 14.5 $\pm$ 5.7 &  0.74 &  126 $\pm$   50 & 19.5 &  17 & 11.3 $\pm$ 4.4 & 5.7 $\pm$ 4.4 &  0.50 &   66 $\pm$   51  \\ 
HB~0850+440 & 0.514 & 121.1 &   435 &  36 & 19.5 $\pm$ 5.7 & 16.5 $\pm$ 5.7 &  0.85 &  149 $\pm$   51 & 20.9 &  27 & 17.0 $\pm$ 5.4 & 10.0 $\pm$ 5.4 &  0.58 &   94 $\pm$   50  \\ 
PG~1001+291 & 0.330 & 131.8 &   382 &  45 & 19.5 $\pm$ 5.7 & 25.5 $\pm$ 5.7 &  1.31 &  198 $\pm$   44 & 19.8 &  33 & 15.2 $\pm$ 5.1 & 17.8 $\pm$ 5.1 &  1.17 &  160 $\pm$   45  \\ 
PG~1358+043 & 0.427 & 130.7 &   434 &  13 & 19.5 $\pm$ 5.7 & -6.5 $\pm$ 5.7 & -0.33 &  -57 $\pm$   50 & 20.4 &  13 & 18.0 $\pm$ 5.5 & -5.0 $\pm$ 5.5 & -0.28 &  -46 $\pm$   50  \\ 
PG~1704+608$^b$ & 0.372 & 131.9 &   409 &  15 & 19.5 $\pm$ 5.7 & -4.5 $\pm$ 5.7 & -0.23 &  -39 $\pm$   49 & 20.1 &  14 & 18.5 $\pm$ 5.6 & -4.5 $\pm$ 5.6 & -0.24 &  -40 $\pm$   50  \\ 
\hline  
Average &  &  &  & 26.5 $\pm$ 11.7 & & 7.0 $\pm$ 2.0\tablenotemark{c} &  0.36 &   57 $\pm$   17\tablenotemark{c} & & 20.8 $\pm$ 9.2 & & 5.3 $\pm$ 1.8\tablenotemark{c} &  0.32 &   48 $\pm$   17\tablenotemark{c}  \\ 
Radio Loud &  &  & & & &  & &   22 $\pm$   25\tablenotemark{c} &  & &  &  & &   23 $\pm$   25\tablenotemark{c}  \\ 
Radio Quiet &  &  & & & & &  &   92 $\pm$   24\tablenotemark{c} &  & &  & &  &   73 $\pm$   24\tablenotemark{c}  \\ 
\enddata  
\tablenotetext{a}{Units of $(h^{-1}_{100} \ \rm Mpc)^{1.77}$. Error calculations are described in text.} 
\tablenotetext{b}{Radio-Loud Quasars} 
\tablenotetext{c}{Error is $\frac{\sqrt{\sum \sigma_i^2}}{n} $} 
\end{deluxetable} 

\begin{deluxetable}{lccc} 
\tablenum{6} 
\tablewidth{0pt} 
\tablecaption{Comparison of Individual \bgq \ Values\tablenotemark{a} 
\label{indivbgq}} 
\tablehead{\colhead{Object} & \colhead{Yee \&} & \colhead{McLure \&} & \colhead{This work} \\ & \colhead{Ellingson 1993}& \colhead{Dunlop 2000} & \colhead{} }  
\startdata  
1004+431    & \phn $-$22 $\pm$ 24 	& \phn $-$16 $\pm$ \phn 71 & \phn $-$23 $\pm$ 44 \\ 
1302$-$102  & \phm{$-$}\phn 62 $\pm$ 52  & \phm{$-$}\phn 94 $\pm$ 107   &\phm{$-$}\phn 41 $\pm$ 43 \\ 
1545+210    & \phm{$-$}129 $\pm$ 66 	& \phm{$-$}113 $\pm$ 109      & \phm{$-$}132 $\pm$ 43 \\ 
1704+608    & \phn $-$58 $\pm$ 41 	& \nodata             & \phn $-$40 $\pm$ 50  \\ 
2349$-$01~    & \phm{$-$}\phn 64 $\pm$ 44  & \phm{$-$}179 $\pm$ 130      & \phm{$-$}\phn 49 $\pm$ 44 \\ 
\enddata
\tablenotetext{a}{\bgq \ in units of ($h_{100}^{-1}$~Mpc)$^{1.77}$}
\end{deluxetable}

\begin{deluxetable}{lrrrrl} 
\tablenum{7} 
\tablewidth{0pt} 
\tablecaption{Comparison of Average Radio-Loud and Quiet \bgq \ Values\tablenotemark{a}
\label{rlrqbgq}} 
\tablehead{\colhead{Reference} & \colhead{Radio-Loud} & \colhead{N\tablenotemark{b}} & \colhead{Radio-Quiet} & \colhead{N\tablenotemark{b}} & \colhead{z} }  
\startdata  
Archive F606W\tablenotemark{c}  & $66 \pm 18 $ & 6 & $57 \pm 12 $ & 13 & $0.155 \le z \le 0.286$\\ 
Archive F702W\tablenotemark{c}  & $23 \pm 25 $ & 4 & $73 \pm 24$ & 4 & $0.223 \le z \le 0.514$\\ 
LBQS  & $-13 \pm 21$ & 6 & $-18 \pm 15$ & 10 & $0.39 < z < 0.504$\\ 
\\
McLure \& Dunlop 2000\tablenotemark{c}  & $ 79\pm 15$ & 10 &  $96 \pm 22$  & 13 & $0.086 \le z \le 0.286 $\\ 
Wold et al. 2000 & $78 \pm 22 $ & 21 &  \nodata & \nodata &$0.5 \le z \le 0.8$ \\ 
Fisher et al. 1996\tablenotemark{c}  & $84~_{-~27}^{+~33}$ & 6 & $72~_{-~19}^{+~20}$ & 14 & $0.086 \le z \le 0.286 $ \\ 
Smith et al. 1995\tablenotemark{d} & \nodata  &\nodata & $20~_{-~~8}^{+~14}$ & 169 & $z < 0.3$\\
Ellingson et al. 1991  & $56 \pm 22$ & 53 & $21 \pm 12$ & 43 & $0.3 < z < 0.6$ \\ 
\enddata
\tablenotetext{a}{\bgq \ in units of ($h_{100}^{-1}$~Mpc)$^{1.77}$}
\tablenotetext{b}{Number of quasars in sample}
\tablenotetext{c}{Archive F606W and Fisher et al. samples use the same $HST$ data
and are identical except for one radio-quiet quasar,
HE~1029$-$1401, which is included in the Fisher et al. sample but not in the Archive F606W sample.  The McLure \& Dunlop sample includes all the Archive F606W quasars 
and one of the Archive F702W quasars, and their analysis is based on the same 
$HST$ data.}
\tablenotetext{d}{Converted published value of ${\langle B_{gq} \rangle}/{\langle B_{gg} \rangle}$ to $\langle B_{gq} \rangle$ \ assuming $\langle B_{gg}\rangle = 20$~(\h~Mpc)$^{1.77}$}
\end{deluxetable}

\end{document}